\documentclass[prb,aps,twocolumn,superscriptaddress,longbibliography,floatfix]{revtex4-2}

\usepackage[
  colorlinks=true,
  urlcolor=blue,
  linkcolor=blue,
  citecolor=blue,
  bookmarks=false,
  pdftitle={Topological nodal line semimetals},
  pdfauthor={Ankur Das}
]{hyperref}

\usepackage{import}
\usepackage{subfiles}
\usepackage{graphicx}
\usepackage{bm}
\usepackage{amsmath}
\usepackage{amssymb}
\usepackage{mathtools}
\usepackage{amsfonts}
\usepackage[all]{xy}
\usepackage{xspace}
\usepackage{accents}
\usepackage{stmaryrd}
\usepackage{tabularx}
\usepackage{extarrows}
\usepackage{float}
\usepackage{multirow}
\usepackage{makecell}
\usepackage[capitalise]{cleveref}
\usepackage[dvipsnames]{xcolor}
\usepackage[normalem]{ulem}
\usepackage{pdfpages}
\makeatletter
\AtBeginDocument{\let\LS@rot\@undefined}
\makeatother

\usepackage{siunitx}

\newcommand{\comment}[1]{}
\newcommand\beq{\begin{equation}}
\newcommand\eeq{\end{equation}}

\begin{document}

\title{Stable nodal line semimetals in the chiral classes in three dimensions}

 \author{Faruk Abdulla}
\affiliation{Harish-Chandra Research Institute, A CI of Homi Bhabha National
Institute, Chhatnag Road, Jhunsi, Prayagraj (Allahabad) 211019, India}
\author{Ganpathy Murthy}
\affiliation{Department of Physics and Astronomy, University of Kentucky, Lexington, KY, USA}
\author{Ankur Das}
\email{ankur.das@weizmann.ac.il}
\affiliation{Department of Condensed Matter Physics, Weizmann
Institute of Science, Rehovot, 76100 Israel}

\begin{abstract}
It has been realized over the past two decades that topological nontriviality can be present not only in insulators but also in gapless semimetals, the most prominent example being Weyl semimetals in three dimensions. Key to topological classification schemes are the three ``internal" symmetries, time reversal ${\cal T}$, charge conjugation ${\cal C}$, and their product, called chiral symmetry ${\cal S}={\cal T}{\cal C}$. In this work, we show that robust topological nodal line semimetal phases occur in $d=3$ in systems whose internal symmetries include ${\cal S}$, without invoking crystalline symmetries other than translations. Since the nodal loop semimetal naturally appears as an intermediate gapless phase between the topological and the trivial insulators, a sufficient condition for the nodal loop phase to exist is that the symmetry class must have a nontrivial topological insulator in $d=3$. Our classification uses the winding number on a loop that links the nodal line. A nonzero winding number on a nodal loop implies robust gapless drumhead states on the surface Brillouin zone. We demonstrate how our classification works in all the nontrivial chiral classes and how it differs from the previous understanding of topologically protected nodal line semimetals.
\end{abstract}

\maketitle

\section{Introduction}

The theoretical understanding of free electron band structure has undergone a revolution in 
recent times, with topology
\cite{TKNN_1982,Haldane1983,KaneMele2005,BernevigHughesZhang2006,FuKane2007,
MooreBalents2007,Roy2009,FuKaneMele2007,HasanKane2010,Moore2010,Bansil_Lin_Das_2016}
playing a central role. In three dimensions, topological insulators
 \cite{MooreBalents2007,FuKaneMele2007,Roy2009,Moore2010,franz2013,Bansil_Lin_Das_2016,
HasanKane2010, Ryu2010}, which necessarily manifest  gapless modes at the boundary  \cite{JackiwRebbi1976, Teo_2010}, were the first examples found of this new paradigm. Shortly thereafter, it was realized that gapless systems in $d=3$ with bulk band crossings
could also be topological \cite{Murakami2007,Wan2011,Burkov_Balents_2011,Burkov2011}.
However, for most of the cases \cite{Murakami2007,Wan2011,Armitage2018,Young2012,Wang2012, Wehling2014,Armitage2018,Burkov2011,Fang2016} (with some exceptions 
\cite{Heikkila2015,Chang2017,das2020,das2022}) we can still use the ideas of topological insulators to classify
these band crossing by enclosing them in a lower dimensional surface in the Brillouin Zone (BZ) such that the Hamiltonian on the lower dimensional surface
is gapped and hence can be thought of as representing a lower dimensional topological
insulator \cite{Ryu2010,ChiuTeoSchnyderRyu2016,Armitage2018}.

For example, in Weyl semimetals (WSMs), a pair of bands cross at an even number of isolated points in the BZ. Either inversion \cite{Murakami2007} or time reversal \cite{Burkov2011} (or both) must be broken in order to have a Weyl semimetal. 
 Each Weyl point is a source of Chern flux on a 2D surface that encloses it in the BZ. Near the
Weyl nodes, the bands are nondegenerate and disperse linearly, giving a density of
states $\rho(E) \sim (E-E_F)^2$.  Weyl semimetals are exceptional in
requiring no symmetry other than translations \cite{Armitage2018}. Dirac semimetals require both
inversion and time-reversal to exist, with the isolated Dirac touchings being four-fold degenerate.
Near the Dirac points, bands are two-fold degenerate and linearly dispersing. Both Weyl and Dirac
semimetals are topological point node semimetals whose band-crossings are zero-dimensional.

The focus of our work is topologically protected nodal line semimetals (TNLSMs) \cite{Burkov2011, Fang2016}. For such systems,
the dispersion around the touching is linear in the two directions perpendicular to 
the nodal line, giving a density of  states $\rho(E) \sim |E-E_F|$. In nexus semimetals, 
nodal lines of a lower degree of degeneracy meet  at a higher degeneracy point at zero 
energy in the Brillouin zone (BZ) \cite{Chang2017,Heikkila2015,das2020,das2022}.

Previous investigations on nodal line semimetals can be divided into two broad categories. One body of work focuses on ``crystalline" symmetries that can protect a topological
nodal line \cite{Chiu2014, Chen2015TopologicalCM,Kim2015SurfaceSO, Yamakage2016, Bian2016, Bian2016n, Sun2017TopologicalNL,  Nie2019, Wang2021TopologicalNC,Fang2015,Kim2015, Weng2015TopologicalNS, Yu2015TopologicalNS, Huang2016,Fang2015,Schoop2016,Hong2018MeasurementOT,Meng2020ANH, Wang2021SpectroscopicEF,Xie2021KramersNL}. By crystalline symmetries, we mean symmetries that are either explicitly symmetries of the crystal other than translations (such as inversion, mirror symmetry, non-symmorphic symmetries, etc.) or $SU(2)$ spin-rotation symmetry.  

Another body of work assumes internal symmetries (${\cal T}$, ${\cal C}$, and ${\cal S}$) only  \cite{Horava_2005,Zhao_Wang_2013,Matsuura_2013,Zhao_Wang_2014}.  In this approach, one classifies the putative topologically protected nodal line as follows: First, the nodal line is enclosed in a lower dimensional surface in the BZ, such that the Hamiltonian restricted to the enclosing surface has the same internal symmetries as the full model and is gapped. For example, if any anti-unitary symmetry is among the internal symmetries, both $\mathbf{k}$ and  $-\mathbf{k}$ must be present on the enclosing surface. We will call such enclosing surfaces centrosymmetric. 
Next, the topological invariant corresponding to the given symmetry class is computed on the enclosing surface. In some symmetry classes, one needs to extend the Hamiltonian to extra dimensions (on which the Hamiltonian must satisfy all the symmetries of the parent class and be gapped) to obtain the topological invariant. If the invariant is nontrivial on the enclosing surface, the nodal line is topologically protected. 

Here, we want to emphasize that the existence of nontrivial indices for a Fermi surface of a given dimension in a given symmetry class \cite{Zhao_Wang_2013,Matsuura_2013} does not guarantee that one can construct a periodic Hamiltonian in that class that realizes this type of Fermi surface. For example, the classification by Refs. \cite{Zhao_Wang_2013,Matsuura_2013} predicts a nontrivial index  for Fermi points in class CII in three dimensions. However, in this work, we explicitly show that there are no  stable Fermi points in class CII in three dimensions. There is only one stable gapless phase, which is  a nodal line semimetal. A similar result holds for class AII in three dimensions as  well (see Appendix \ref{App:AII}).

The goal of this work (and the accompanying short paper \cite{MainPaper}) is to provide a different and simpler classification of TNLSMs in chiral classes based on the winding number of a test loop that links the nodal line. A sufficient condition for the given chiral class to have a TNLSM phase is that the class should have a nontrivial topological insulator phase in $d=3$ \cite{Altland1997,Heinzner2005,Schnyder2008}. We are not sure this condition is necessary as well, because of the well-known 2D counterexample of spinless graphene (class BDI, trivial in 2D) having topologically protected Dirac nodes. Of all the chiral classes, only class BDI is topologically trivial in 3D, and will be ignored henceforth. The rest of the chiral classes (AIII, CI, CII, and DIII) all have a generic and robust gapless phase near the transition between the topological and trivial insulators. By generic, we mean that all couplings allowed by the given symmetry class are present in the Hamiltonian. It is straightforward to show by contradiction that this gapless phase cannot be a Weyl semimetal: Assume that it is indeed a WSM. Then, if one encloses the putative Weyl point in a generic two-dimensional surface in the BZ (which does not necessarily satisfy the symmetries of the parent class), the Hamiltonian restricted to this surface is in class AIII, which has a trivial topology in $d=2$. Hence the gapless phase cannot generically be a Weyl semimetal. Therefore the phase must be a nodal line semimetal.

The key difference in our approach to classifying TNLSMs as compared to the previous one \cite{Zhao_Wang_2013,Matsuura_2013,Zhao_Wang_2014} is that the loop enclosing the nodal line need not be chosen in a centrosymmetric way, that is, both ${\bf k}$ and $-{\bf k}$ need not be present on the test loop if the symmetry class has an anti-unitary symmetry.  This means that the Hamiltonian restricted to the enclosing loop only has chiral symmetry, and is in class AIII. Since this class has a nontrivial topology in $d=1$, this enables us to characterize TNLSMs in chiral classes in $d=3$ by the  winding number $W$ (henceforth called  the topological charge of the nodal loop) \cite{Ryu2010,ChiuTeoSchnyderRyu2016}  of the Hamiltonian restricted to the test loop, 
\begin{align}\label{eq:invariant}
    W = \frac{1}{2\pi i} \int_{0}^{2\pi} dk ~ \textrm{Tr}\left(Q^{-1} \partial_k Q\right).
\end{align}
In this expression, the Hamiltonian has been brought to block off-diagonal form (as can always be done in a chiral class), and the matrix $Q(k$) 
in \cref{eq:invariant} is the upper right block. 

Clearly, the sign of the winding number depends on the sense of the enclosing loop. For classes where time reversal is present, nodal loops generically appear in pairs. Choosing the senses of the enclosing loops in the two nodal loops in a well-defined way, one obtains a relation between the winding numbers of nodal loops related by time reversal. We will elaborate on this later.

For classes AIII, CI, and DIII, our approach gives  essentially the same classification as the previous approach \cite{Zhao_Wang_2013,Matsuura_2013,Zhao_Wang_2014}, with the invariant associated with the topological semimetal being classified by the integers. The only difference is that our loops do not have to be centrosymmetric. For class CII, however, the classifications appear to be different. In our approach, the topology of a given nodal loop is simply the winding number on an enclosing loop that encloses only that nodal loop, which belongs to the integers. The previous approach says that nodal loop semimetals in class CII are classified by $Z_2$. In this case, the invariant has to be defined on a centrosymmetric loop extended into two extra dimensions, forming a 3-Torus, with the Hamiltonian being gapped everywhere on the 3-Torus, and reducing to a topologically trivial model at a given point on the 3-Torus. In order to determine whether a given model is trivial or nontrivial, in principle, one has to compute the 3-Torus invariant on all possible centrosymmetric starting loops, which is a difficult task, one which we have not been able to perform. What we have been able to do is to take a particularly simple enclosing loop (the $k_z$ axis of the 3D BZ) in a model which we know has nontrivial nodal loops. The 3-Torus invariant happens to be trivial on that simple enclosing loop (c.f. \cref{appsec:Z_2}). Unfortunately, so does the winding number of our approach, so no conclusions can be drawn from these calculations. The precise relation between our approach and that of Refs. \cite{Zhao_Wang_2013,Matsuura_2013,Zhao_Wang_2014} remains unclear to us.

Going further, we how our winding number predicts the presence and degeneracies of the zero-energy drumhead modes \cite{Heikkil_2011, Burkov2011, Kim2015, Weng2015TopologicalNS,
Yu2015TopologicalNS, Chen2015TopologicalCM, Kim2015SurfaceSO, Hong2018MeasurementOT}
on an open surface of the TNLSM. In class CII which has ${\cal T}^2={\cal C}^2$, nodal loops related by time reversal have the same winding number. Their projections onto the surface BZ of a given open surface in real space may overlap either partially or fully, or not overlap at all. In the region of overlap, there are two-fold degenerate drumhead modes, while in the non-overlapping region, the drumhead modes are singly degenerate. In classes CI and DIII, which have ${\cal T}^2=-{\cal C}^2$, each pair of nodal loops related by time reversal have opposite winding numbers. In this case, if the projections of the two nodal loops overlap on the surface BZ of an open surface in real space, there are no gapless drumhead states in the region of overlap. However, in regions of the surface BZ where the projections of the nodal loops do not overlap, gapless drumhead states are recovered. Thus, not only is the winding number of a loop a powerful classification tool in the bulk, the algebraic sum of the winding numbers of all the loops that project to a point on the surface BZ also predicts the presence and degeneracy of the gapless drumhead modes on the corresponding open surface.

The plan of the paper is as follows: Since the simple but illuminating case of class AIII is discussed
in Ref. \onlinecite{MainPaper}, we refrain from repeating that here. 
We start with  class CII in (\cref{Sec:CII}), where we set up and fully explore a minimal model with an $8\times8$ Hamiltonian and show how the generic gapless phase with two-fold degenerate nodal loops is obtained. We know that the model is in a topologically protected nodal loop semimetal because the winding numbers of the nodal loops are nonzero. In (\cref{Sec:CI}) and (\cref{Sec:DIII}) we similarly explore classes CI and DIII. The next section, \cref{Sec:Surface}, turns to the consequences of our classification for surface
states in TNLSMs. We end by summarizing our findings and presenting an outlook for 
future prospects in \cref{Sec:Summary}. Two appendices present some subsidiary calculations. In \cref{appsec:Z_2}, we compute the topological invariant of the previous approach \cite{Zhao_Wang_2013,Matsuura_2013,Zhao_Wang_2014} on a specific loop traversing the BZ, and show that it is trivial. In \cref{App:AII} we discuss class AII in three dimension that has gapless nodal points (Weyl semimetals).

\section{Class CII}

\label{Sec:CII}

A Hamiltonian in symmetry class CII must have time-reversal
(${\cal T}$), particle-hole (${\cal C}$) and the chiral (${\cal S}$) symmetry with ${\cal T}^2=-1, {\cal C}^2=-1$. In $d=3$ a topologically nontrivial phase classified by $Z_2$ exists in this class. A minimal Hamiltonian with spin, orbital, and chiral (sublattice) degrees of freedom that describes the transition
between 3D topological and trivial insulating states is 
\begin{align}
    H_0({\bf k}) = \sum_{i=1}^{3} \eta^x \otimes \left( k_i  \gamma^i + m \gamma^4\right).
\label{eq:HCII_min}
\end{align}
We denote the Pauli matrices for the sublattice/chiral space by $\eta^a$, those in the 
orbital space by $\tau^a$, and those in the spin space by $\sigma^a$. Our $4\times4$ $\gamma$ matrices
are represented in the spin and orbital basis as
$\gamma \equiv \tau \otimes \sigma $. Our chosen representation is the following:
$\gamma^1=\tau^z \sigma^x, ~  \gamma^2=\tau^z \sigma^y, ~\gamma^3=\tau^y \sigma^0,
~\gamma^4=\tau^x\sigma^0$. The rest of the gamma matrices are
$\gamma^5=\gamma^1\gamma^2\gamma^3\gamma^4=\tau^z\sigma^z$ and $\gamma_{ij}=-\frac{i}{2}[\gamma^i,
\gamma^j]$ for $i<j=1,2,\dots, 5$.
The time-reversal and particle-hole operations are realized by ${\cal T}=-i\sigma^y K$
and ${\cal C}=i\eta^z\sigma^y K$, satisfying ${\cal T}^2=-1$,
${\cal C}^2=-1$.
The chirality operator in this representation is ${\cal S} = {\cal T}{\cal C} =
\eta^z$. The Hamiltonian also has an accidental  inversion
symmetry represented by ${\cal I}=\eta^x\tau^x$. The transition
from topological to trivial insulating phase occurs through a Dirac point at the
origin ${\bf k}=0$. We have also taken the continuum limit, implicitly assuming that all the interesting physics occurs in a small region near the $\Gamma$ point ${\bf k}=0$. We note that this assumption is inadequate when we go on to compute the topological invariant of the previous approach; we will have to extend the model to a Hamiltonian function of ${\bf k}$ periodic in the BZ.  In the following, we will show that the transition from the 3D
topological to trivial insulating states in class CII occurs through a gapless
phase which  is generically a nodal line semimetal, which is topologically protected against getting gapped by the nontrivial winding numbers of the nodal loops.

\begin{table}[ht]
  \begin{center}
    \begin{tabular}{p{40mm} p{20mm} p{8mm}} 
      \hline 
      \hline
      \textbf{Operators} & \textbf{TR} & \textbf{I}\\ [1mm]
      \hline
      $\eta^x \otimes {\bf a}$ ~~ $(\Gamma_1, \Gamma_2, \Gamma_3) $ & $-1$ & $-1$ \\ [1mm]
      $\eta^x \otimes \gamma^4$ ~ $(\Gamma_4)$ & $+1$ & $+1$ \\ [1mm]
      $\eta^x \otimes \gamma^5$ ~ $(\Gamma_5)$ & $-1$ & $-1$  \\ [1mm]
      $\eta^x \otimes \gamma^0$ & $+1$ & $+1$ \\ [1mm]
      $\eta^x \otimes \epsilon$ & $+1$ & $-1$ \\ [1mm]
      $\eta^x \otimes {\bf p} $ & $+1$ & $-1$ \\ [1mm]
      $\eta^x \otimes {\bf b} $ & $-1$ & $+1$ \\ [1mm]
      $\eta^x \otimes {\bf b}'$ & $-1$ & $+1$ \\ [3mm]

      $\eta^y \otimes {\bf a} $ & $+1$ & $+1$ \\ [1mm]
      $\eta^y \otimes \gamma^4$ & $-1$ & $-1$  \\ [1mm]
      $\eta^y \otimes \gamma^5$ & $+1$ & $+1$  \\ [1mm]
      $\eta^y \otimes \gamma^0$ ~ $(\Gamma_6)$ & $-1$ & $-1$  \\ [1mm]
      $\eta^y \otimes \epsilon$ & $-1$ & $+1$  \\ [1mm]
      $\eta^y \otimes {\bf p} $ & $-1$ & $+1$ \\ [1mm]
      $\eta^y \otimes {\bf b} $ & $+1$ & $-1$ \\ [1mm]
      $\eta^y \otimes {\bf b}'$ & $+1$ & $-1$ \\ [1mm]
      \hline 
    \end{tabular}
    \caption{Transformation properties of Dirac matrices (representation given in \cref{eq:HCII_min})
    which anticommute with ${\cal S} = \eta^z$.
    For eight by eight Gamma matrices, there are seven Gamma matrices which anticommute with each 
    other. Six of the seven anticommuting Gamma matrices are given in the table. The last one $\Gamma_7 = (-i)
    \prod_{i=1}^6 \Gamma_i = \eta^z$,  which commutes with  ${\cal S} = \eta^z$, breaks the chiral symmetry.      \label{tab:table1}
}
  \end{center}
\end{table}

The Hamiltonian in  \cref{eq:HCII_min} is not the most general 
Hamiltonian in class CII because there are many terms that can be added to
$H_0({\bf k})$ without breaking  ${\cal T}$ and ${\cal C}$. One can add any linear combination of time reversal even $\gamma$'s times an $\eta^x$, and any linear combination of time reversal odd $\gamma$'s times an $\eta^y$. 
The most general (minimal dimensional)  Hamiltonian in class CII is 
\begin{equation}
\begin{aligned}
    H_{\textrm{CII}}({\bf k}) =&  H_0({\bf k}) + \mu \eta^x \otimes \gamma^0 + \eta^y\otimes{\bf A}_a\cdot{\bf a}\\
    &+A_5\eta^y\otimes\gamma^5+ 
    \eta^y \otimes {\bf V}_{b} \cdot {\bf b}  +  \eta^y \otimes 
    {\bf V}_{b'} \cdot {\bf b}'\\
    &+ \eta^x \otimes {\bf V}_{p} \cdot {\bf p} +  V_{\epsilon} \eta^x \otimes \epsilon,
\end{aligned}
\label{eq:HCII}
\end{equation}
where the four by four  ${\bf a}$, ${\bf b}$, ${\bf b}'$, ${\bf p}$ and $\epsilon$  matrices  are ${\bf a}=\left(\gamma^1,\gamma^2,\gamma^3 \right)$, ${\bf b} =\left(\gamma_{23}, \gamma_{31}, \gamma_{12}\right)$,  ${\bf b}'= \left(\gamma_{15}, \gamma_{25}, \gamma_{35} \right)$, ${\bf p}=\left(\gamma_{14}, \gamma_{24}, \gamma_{34}\right)$ and $\epsilon=\gamma_{45}$ respectively, and $\gamma^0=\tau^0 \sigma^0$. The transformation properties of the various classes of operators under ${\cal T},\ {\cal C}$, and ${\cal I}$ are summarized in \cref{tab:table1}. The inversion preserving terms $m$, $\mu$, $A$'s, as well as the inversion broken terms $V$'s can be any even function of the momentum components. Without loss of generality, we assume the following quadratic dependence on momenta near the Gamma point: $ X_{\alpha}({\bf k}) = X_{\alpha 0} - \sum_{ij} t^{(X_\alpha)}_{ij} k_i k_j$, where $X_{\alpha}$ is an element of the set $X = \{m, \mu, {\bf A}_a, A_5, {\bf V}_b, {\bf V}_b', {\bf V}_p, V_{\epsilon} \}$, and all $X_{\alpha0}$, $t^{(X_\alpha)}_{ij}$ are real numbers. We will assume the matrices $t_{ij}$ to be positive definite because the resulting nodal loops remain confined to small $k$, where our continuum Hamiltonian is valid. 

Diagonalizing the Hamiltonian of \cref{eq:HCII} in its most general form and finding the locus of its gapless points is a difficult task. To make this task easier, we will add a selected sequence of perturbations to the minimal Hamiltonian of \cref{eq:HCII_min}, taking two different approaches.  In the first approach, we  start with perturbations which preserve the accidental inversion symmetry. This leads to four-fold degenerate nodal loops. We then add perturbations that break inversion to obtain a pair of two-fold degenerate nodal loops. Finally, we show that these two-fold degenerate nodal loops are generic and stable to all symmetry-allowed perturbations. In the second approach, we first show that adding a single inversion-breaking perturbation will result in four  four-fold degenerate Dirac points. This occurs because  when a single perturbation breaking the standard inversion ${\cal I}=\eta^x\tau^x$ is added one can define a modified inversion operator under which the Hamiltonian is symmetric. We then add further perturbations that break the modified inversion ($\eta^x {\bf V}_p\cdot{\bf p}$ and $\eta^y {\bf V}_b\cdot {\bf b}$) and show that the locus of gapless points describes  pairs of two-fold degenerate nodal loops with nontrivial winding numbers. Once we have nodal loops with nontrivial winding number, we know that they are stable against small deformations, and are thus generic and robust. 

\subsection{\bf Inversion-preserving perturbations:}
We recall from \cref{tab:table1} that the $V$ terms in \cref{eq:HCII} are inversion breaking while the other perturbations preserve inversion. Thus, the most general inversion-preserving Hamiltonian in class CII is
\begin{align}
H^{IP}_{\text{CII}}({\bf k}) = H_0({\bf k}) + \mu \eta^x \otimes \gamma^0  +A_5\eta^y\otimes\gamma^5 
+ \eta^y\otimes{\bf A}_a\cdot{\bf a}.
\label{eq:HCIIInv}
\end{align}
Since the Hamiltonian $H^{IP}_{\text{CII}}({\bf k})$ has time-reversal and inversion
symmetry, every band must be doubly degenerate for all ${\bf k}$, leading to four distinct eigenvalues. The additional ${\cal I}{\cal T}$
symmetry allows us to diagonalize the Hamiltonian by squaring, rearranging, and squaring again. 
We find the following energy spectrum.
\begin{equation}
    E^2({\bf k}) = \left({\bf k}^2 + m^2 + \tilde{m}^2\right) \pm 2 \sqrt{\tilde{m}^2({\bf k}^2 + m^2)
    - \left({\bf k} \cdot {\bf A}_a\right)^2 }
\end{equation}
where, $\tilde{m}^2= \mu^2 + A_5^2 + {\bf A}_a^2$.
The condition for zero-energy solutions is easily obtained. 
\begin{equation}\label{eq:4degNL}
    ({\bf k}^2+m^2- \tilde{m}^2)^2 + 4\left({\bf k}\cdot {\bf A}_a\right)^2 = 0. 
\end{equation}
Evidently, the solution space is given by the intersection of the two surfaces 
${\bf k}^2+m^2- \tilde{m}^2 = 0$  and ${\bf k}\cdot {\bf A}_a=0$. Since two surfaces in $d=3$, if they intersect at all,  generically intersect in a line, the gapless solutions form a  nodal
loop. Since it represents the intersection of two two-fold degenerate branches of the spectrum, the nodal loop is four-fold degenerate. Note that for any small deformations in
the parameters in the Hamiltonian of \cref{eq:HCIIInv} will slightly deform the two intersecting surfaces, and thus slightly deform the nodal loop. The only way to get rid of the nodal loop is to shrink it to zero and then gap it out. So the gapless phase with a four-fold degenerate nodal loop is stable  to arbitrary inversion-preserving perturbations. Computing  the topological 
charge $W$ of the nodal loop, and we find that it carries winding number $W=2$, making it 
topologically nontrivial.

\subsection{\bf Four-fold degenerate nodal line to two-fold degenerate nodal lines}

The obvious next step is to break inversion and examine what happens to the four-fold degenerate nodal loop. Clearly, inversion-breaking perturbations 
cannot gap out the nodal loop because it carries a nontrivial winding number. As we will see, the inversion-breaking perturbations  
 transform the four-fold degenerate nodal loop into a pair of two-fold degenerate
nodal loops related by time reversal,  each carrying an identical topological charge of $W=1$.

\begin{figure}[h]
\centering
\includegraphics[width=0.95\columnwidth]{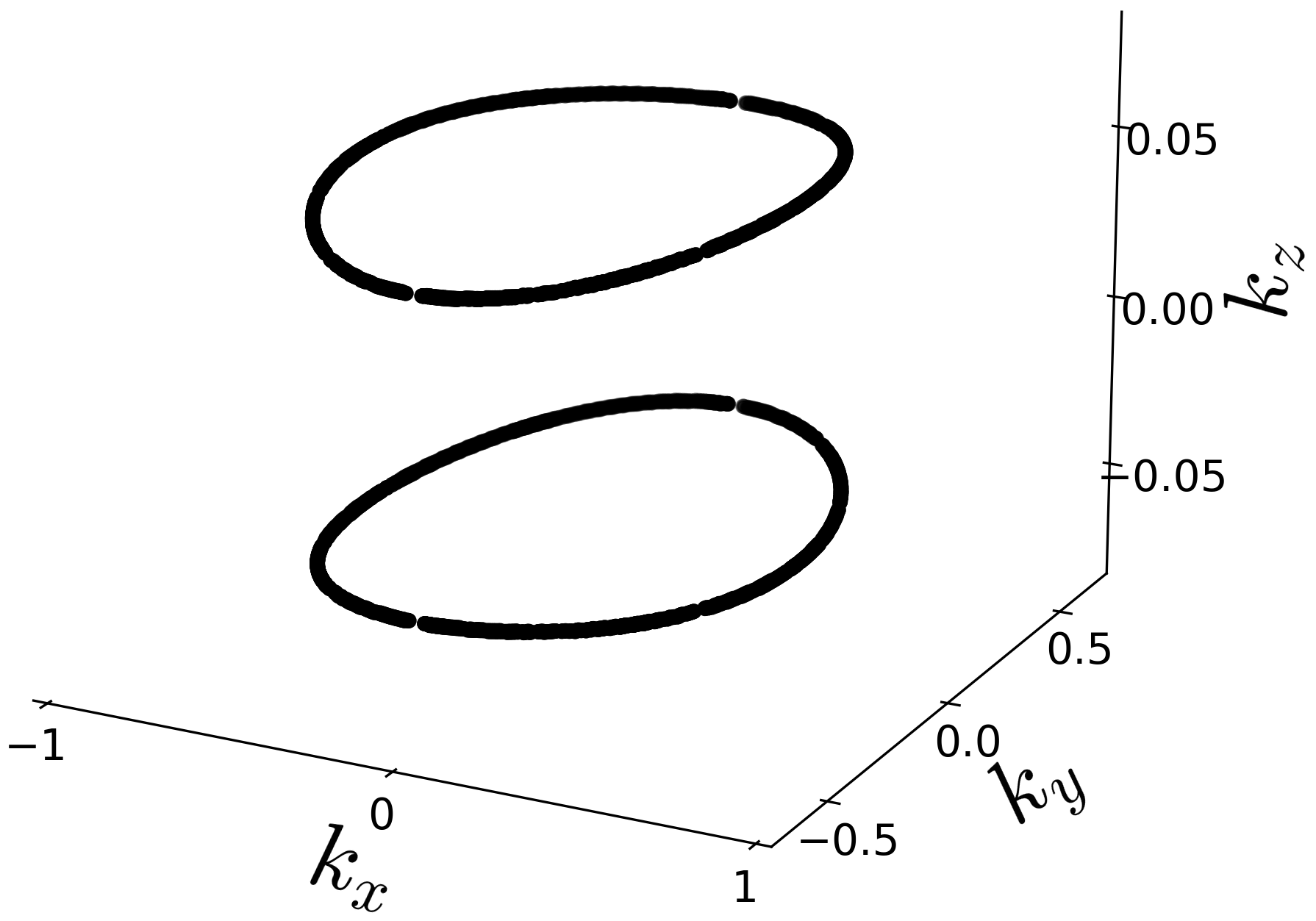}
\caption{Nodal loop solution of the CII model whose gapless condition is given in
\cref{eq:CIIfinalCond}. The parameters are ${\bf A}_a=(0, 0, 0.7)$, $V_{\epsilon}=0.1$, and $m({\bf k}) = m_0 - \sum_{i=1}^{3}t_i k_i^2$, where $m_0=0.4$ and $t_{1}=0.1$,
$t_{2}=0.2$, $t_{3}=0.3$. The two nodal loops are time-reversal partners of each other
carrying the same winding number $W=1$.}
\label{fig:CIILoop}
\end{figure}
 
To implement this, we add $V_{\epsilon}$ to
$H^{IP}_{\text{CII}}({\bf k})$ and solve for its gapless solutions explicitly. 
For simplicity, we will set 
$\mu=0$, $A_5=0$ and use ${\bf A}_a = (0, 0, A)$. Note that with this choice, the 
four-fold degenerate nodal loop solution of \cref{eq:4degNL} will lie in the 
$k_x$-$k_y$ plane at $k_z=0$. Thus,  with  the inversion-breaking  term $V_{\epsilon}
\eta^x \otimes \epsilon$, we have the following Hamiltonian
\begin{align}\label{eq:HCIIVe}
H_{\text{CII}, \epsilon} =H_0({\bf k})  + A \eta^y\otimes a_3 + V_{\epsilon} \eta^x \otimes \epsilon. 
\end{align}
Upon squaring, rearranging, and squaring again, we obtain the energies to be 
\begin{align}
    E^2 =& \left({\bf k}^2  +m^2+A^2+V_{\epsilon}^2 \right)  \nonumber\\
    &\pm2\sqrt{(A^2+V_{\epsilon}^2)(k_x^2+k_y^2)+(Am\pm V_{\epsilon}k_z)^2}.
\end{align}
Evidently, all eight eigenvalues are distinct. Zero-energy solutions require
\begin{equation}
    ({\bf k}^2+m^2-A^2-V_{\epsilon}^2)^2=-4(Ak_z\pm V_{\epsilon}m)^2,
\end{equation}
which implies two conditions:
\begin{equation}
    {\bf k}^2+m^2=A^2+V_{\epsilon}^2;\ \ \ Ak_z=\pm V_{\epsilon}m.
\label{eq:CIIfinalCond}
\end{equation}
The four-fold degenerate present at $V_{\epsilon}=0$ splits into a
pair of two-fold degenerate nodal loops related to each other by time reversal symmetry,
one for each solution of the second equation above. By making either $m$ or $V_{\epsilon}$
a generic even function of ${\bf k}$, we break all the lattice symmetries. The
resulting nodal loops will not lie in a plane. As an illustration, the  nodal loops are presented in 
\cref{fig:CIILoop} for a specific choice of parameters. Each of the individual
nodal loops carries a topological charge $W=1$. As mentioned before, in class CII the winding number of two loops related by ${\cal T}$ are identical. This can be traced back to the fact that ${\cal T}^2={\cal C}^2$ in class CII. 

Now that all the lattice symmetries have been broken, it is clear that perturbing the Hamiltonian with other terms from \cref{eq:HCII} will not destabilize the nodal loops, because each individual loop is topologically protected by its nonzero winding number.  The only generic way to remove the pair of loops is to shrink them to points. Thus, we have shown one of our principal claims, that there is a robust gapless phase in class CII, and it is described by pairs of two-fold degenerate nodal loops related by ${\cal T}$. 

One can further envisage breaking ${\hat T}$ in our class CII Hamiltonian, thus converting it to a class AIII Hamiltonian. Since we have already established in Ref. \onlinecite{MainPaper} that generic gapless phases in class    AIII should be TNLSMs, we do not pursue this further.

\begin{figure}[ht]
     \centering
     \includegraphics[width=1\columnwidth]{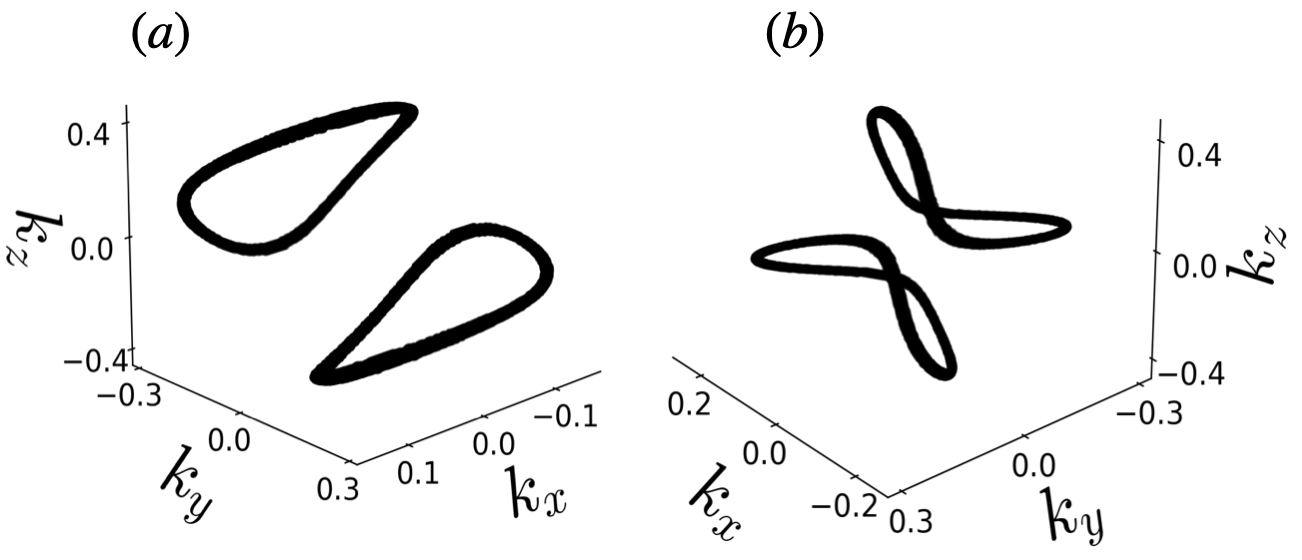}
     \caption{The gapless conditions in \cref{eq:Hsolution} are solved for the two
     different choices of ${\bf V}_b$ for a fixed  ${\bf V}_p = (0.3, 0, 0)$, 
     $m_0=0.1$ and   $t_{1}=1.4, t_{2}=1.2, t_{3}=0.6$. $(a)$ There exist two nodal loops
     for the choice ${\bf V}_b = (0, 0.1, 0.1)$. $(b)$ For the other choice ${\bf V}_b = 
     (0.1, 0.2, 0.2)$, again, there exist two nodal loops. The two nodal loops are
     time-reversal partner of each other carrying the same winding number $W=1$.}
 \label{fig:NL in case 2}
 \end{figure}

\vspace{0.4cm}

\subsection{Four-fold degenerate Dirac points to two-fold degenerate nodal lines}

There is another route to go  from the eight-fold degenerate Dirac point (the massless point of \cref{eq:HCII_min}) to the generic pair of two-fold degenerate nodal loops. The eight-fold degenerate Dirac point first goes to four  four-fold degenerate Dirac points upon adding a single inversion-breaking perturbation. This can occur because there are several different inversion operations, and a single perturbation does not break all of them. When one adds a second inversion-breaking perturbation, all inversions are broken, and the four Dirac points go to four two-fold degenerate nodal loops. As the second inversion-breaking parameter is increased, the four nodal loops coalesce into two nodal loops related by time reversal. For the sake of completeness, we illustrate this route to the generic pair of nodal loops as well.

Let us explicitly show that adding only a single inversion-breaking $V$-type perturbation to \cref{eq:HCII_min}  takes the eight-fold degenerate Dirac point to four four-fold degenerate Dirac points protected by a modified inversion. Suppose one adds only $V_{\epsilon} \eta^x\tau^y\sigma^z$. The Hamiltonian is invariant under ${\cal T}{\cal I}$, which is an antiunitary inversion-like operator. Similarly, if one adds only $V_{p}\eta^x p_z= V_{p}\eta^x\tau^z$, the Hamiltonian is still symmetric under the modified inversion operator ${\cal I}'=\eta^y\tau^y\sigma^z$. Let us be even more explicit. We add only the $\eta^x p_z$ perturbation and start with the Hamiltonian
\begin{align}
    H_p({\bf k})=\eta^x\otimes \big(D({\bf k})+V_p \tau^z\big)
    \label{eq:HCIIp}
\end{align}
Squaring the Hamiltonian we obtain
\begin{align}
    \big(H_p({\bf k})\big)^2={\bf k}^2+m^2+V_p^2+2V_p(k_x\sigma^x+k_y\sigma^y)
\end{align}
The last term has eigenvalues $\pm 2V_p k_{\perp}$, where $k_{\perp}^2=k_x^2+k_y^2$. Therefore, the eigenvalues are 
\begin{align}
    E^2=k_z^2+m^2+(k_{\perp}\pm V_p)^2
\end{align}
The conditions for gapless points to exist are $k_z=0$, $m=0$, and $k_{\perp}=V_p$. The first means that the gapless points lie in the $xy$-plane. Both the second and the third conditions have loop solutions (recall that both $M$ and $V_p$ can have quadratic terms in the components of $k$). Two generic loops in the $xy$-plane, each of  which is symmetric under ${\bf k}\to-{\bf k}$, intersect at four points. Thus, there are generically four four-fold degenerate Dirac points that descend from the eight-fold degenerate Dirac point of $H_0$. 

The next step is to break the accidental ${\cal I}'$ symmetry, implemented by adding a further ${\bf V}_{b}$ perturbation to \cref{eq:HCIIp}. The new Hamiltonian, not possessing any inversion-like symmetries, is given by
\begin{align}
H({\bf k}) = \eta^x \otimes D({\bf k}) + \eta^x \otimes {\bf V}_{p} \cdot {\bf p} +  \eta^y \otimes {\bf V}_{b} \cdot {\bf b}. 
\label{eq:Hpb}
\end{align}
We expect to find eight distinct eigenvalues. While we can diagonalize the Hamiltonian of \cref{eq:Hpb} analytically by solving a quartic, our goal of finding the locus of gapless points is much simpler to achieve.  We square the Hamiltonian and collect terms to obtain
\begin{align}
H^2({\bf k}) = \left({\bf k}^2 + m^2({\bf k})  + {\bf V}_p^2 + {\bf V}_b^2 \right) + 2 h_1({\bf k}),
\label{eq:Hpbsq} 
\end{align}
where the auxiliary Hamiltonian $h_1$ is defined as   
\begin{align}
h_1({\bf k})  =& \eta^z \otimes {\bf a} \cdot {\bf V}_b \times {\bf k} + \eta^z \otimes {\bf p} \cdot {\bf V}_b \times {\bf V}_p\nonumber\\ 
&+ \eta^0 \otimes {\bf b'} \cdot {\bf k} \times {\bf V}_p. 
\label{eq:h1}
\end{align}
Evidently we must find the spectrum of  $h_1({\bf k})$. We square it and
collect terms to obtain
\begin{widetext}
\begin{align} \label{eq:h1sq}
(h_1({\bf k}))^2 = &  \left(|{\bf k} \times {\bf V}_p|^2 +  |{\bf V}_b \times {\bf k}|^2 + 
|{\bf V}_b \times {\bf V}_p|^2  \right) + 2 \left({\bf k} \cdot {\bf V}_p \times {\bf V}_b\right)~ 
h_2({\bf k}) = f_1 +  f_2~  h_2({\bf k}). 
\end{align} 
\end{widetext}
Here $f_1 = \left(|{\bf k} \times {\bf V}_p|^2 +  |{\bf V}_b \times {\bf k}|^2 + |{\bf V}_b \times {\bf V}_p|^2  \right)$, 
$f_2 = 2 \left({\bf k} \cdot {\bf V}_p \times {\bf V}_b\right)$ and 
\begin{align}
 h_2({\bf k}) =   \eta^z \otimes {\bf a} \cdot {\bf V}_p -  \eta^z \otimes {\bf p} \cdot {\bf k} +  \eta^0 \otimes {\bf b'} 
  \cdot {\bf V}_b. 
\label{eq:h2}
\end{align}
We note from \cref{eq:h1sq} that if $ {\bf V}_p \times {\bf V}_b=0$, then
$(h_1({\bf k}))^2$ will be a multiple of the identity matrix  and hence every band of $H({\bf k})$ will 
be doubly degenerate. Since we are looking for the case when all accidental symmetries are broken, we assume $ {\bf V}_p \times {\bf V}_b \neq 0$.
Squaring $ h_2({\bf k})$ and collecting terms as usual we obtain the 
following simple relation 
\begin{align}\label{eq:h2sq}
(h_2({\bf k}))^2 = ( {\bf k}^2  + {\bf V}_p^2  + {\bf V}_b^2)  + 2~h_1({\bf k}) = f_3 +  2~h_1({\bf k})                      
\end{align}
where $ f_3 = ( {\bf k}^2  + {\bf V}_p^2  + {\bf V}_b^2)$. Now combining
\cref{eq:Hpbsq,eq:h2sq}, we find the following useful relation
\begin{align}
H^2({\bf k}) = m^2({\bf k}) + (h_2({\bf k}))^2. 
\label{eq:Hsqh2sq}
\end{align}

\noindent
The above matrix equation implies the following eigenvalue equation
\begin{align}
E^2({\bf k}) = m^2({\bf k}) + \lambda^2({\bf k}),
\label{eq:Esq}
\end{align}
where   $\lambda^2({\bf k})$ is the eigenvalue of the Hermitian matrix $(h_2({\bf k}))^2$. Therefore for  
zero energy (gapless) solutions, we must have (i) $ m({\bf k}) = 0$ and (ii)  $\lambda({\bf k}) = 0$. To find $\lambda({\bf k})$, 
we combine \cref{eq:h1sq,eq:h2sq} to get the following eigenvalue equation 
\begin{align}
\left(\lambda^2({\bf k}) - f_3\right)^2 = 4\left(f_1 + f_2 ~ \lambda({\bf k})\right). 
\label{eq:lambdak}
\end{align}
This is the quartic whose solution enables us to find all the eigenvalues of the Hamiltonian. To achieve our simpler goal, we set $\lambda({\bf k})=0$ in the above equation to obtain the following condition.  
 \begin{align}
 f_3^2 = 4f_1. 
 \end{align}
 Therefore for gapless solutions, the variables  ${\bf k}$, ${\bf V}_p$, ${\bf V}_b$ and $m({\bf k})= 
\left(m_0 - \sum_{i=1}^{3} t_{i}k_i^2\right)$ must satisfy the following  two conditions
\begin{subequations}
\label{eq:Hsolution}
\begin{align}
&  ( {\bf k}^2  + {\bf V}_p^2  + {\bf V}_b^2)^2 = 4\left(|{\bf k} \times {\bf V}_p|^2 +  |{\bf V}_b \times {\bf k}|^2 + 
|{\bf V}_b \times {\bf V}_p|^2 \right) \\
& m({\bf k}) = 0
\end{align}
\end{subequations}
The solution space is the intersection of two surfaces in ${\bf k}$ space, which will generically be a nodal line. It could happen, for a given set of coupling constants, that there are no solutions to the above equations. In that case, we are in one of the insulating phases. We can say with definiteness that if solutions exist for the given set of coupling constants, they must generically be in the form of nodal loops. To show that solutions do exist, we proceed as follows: We note that the first equation depends on $(m_0, t_{ij})$ but is independent of $({\bf V}_p, {\bf V}_b)$, while the second equation depends on $({\bf V}_p, {\bf V}_b)$  but is independent of $(m_0, t_{ij})$. Thus, fixing, say, $(m_0, t_{ij})$, we fix one of the surfaces. We now vary the second surface  by changing the parameters $({\bf V}_p, {\bf V}_b)$ leaving the first surface unaffected. This implies that we can always find choices of parameters that    have nodal loop solutions.  Examples of such solutions for two choices of parameters are shown in \cref{fig:NL in case 2}. We have computed their topological charge and find that each loop carries $W=1$, the two loops being related by time reversal. 

To summarize, by setting ${\bf V}_b=0$ in \cref{eq:Hsolution}, we find that the resulting gapless  conditions describe four Dirac points. Now, if we turn on a tiny ${\bf V}_b$, each four-fold degenerate Dirac point immediately goes to  a two-fold degenerate nodal line as described by the conditions in \cref{eq:Hsolution}. As $|{\bf V}_b|$ increases, the four nodal loops coalesce into a pair of nodal loops related by time reversal. The situation is the same if one starts with ${\bf V}_p=0$ but with a nonzero ${\bf V}_b$, with ${\bf V}_p$ and ${\bf V}_b$ trading places. Pairs of nodal loops related by time reversal exist and are stable to deformations in the generic band structure in the gapless phase.

Previous work on  nodal loop semimetals in class CII \cite{Zhao_Wang_2013,Matsuura_2013,Zhao_Wang_2014} results in a topological charge taking values in   $\mathbb{Z}_2$. We discuss the highly nontrivial calculation of this $\mathbb{Z}_2$ invariant for a particular set of simple centrosymmetric loops in \cref{appsec:Z_2}.

\section{Class CI}
\label{Sec:CI}
\noindent
Hamiltonians belonging to class CI must have time-reversal (${\cal T}$),
particle-hole (${\cal C}$), and the chiral (${\cal S}$) symmetry with 
${\cal T}^2=1, {\cal C}^2=-1$. in $d=3$, nontrivial topological insulators do exist in class CI, and are classified by even integers. A minimal model that describes the transition
between 3D topological and trivial insulating states and also satisfies the above
symmetries is an eight component Dirac Hamiltonian
\begin{align}\label{eq:H0CI}
    H_0({\bf k}) = \eta^x ~{\bf k}.{\bf a} + m\eta^y,
\end{align}
where, ${\bf a} = \left(\tau^y \sigma^z, \tau^y \sigma^x, \sigma^y \right)$.
Here $\sigma$’s, $\tau$'s, and $\eta$’s act 
on spin, orbital, and chiral/sub-lattice space, respectively. The time-reversal and 
particle-hole operations are realized by ${\cal T}=-i\eta^x K$  and ${\cal C}= -\eta^y K$
respectively. They clearly satisfy ${\cal T}^2=1, {\cal C}^2=-1$, confirming that the
Hamiltonian $H_0({\bf k})$ belongs to class CI. The chiral symmetry operation is 
simply given by the product of particle-hole and time-reversal ${\cal S} = {\cal T}{\cal C}
= \eta^z$. Evidently, the transition from topological to trivial insulating phase occurs
through a Dirac point at the origin ${\bf k}=0$. Shortly we will see that this Dirac
point is not stable; rather, it will immediately go to nodal loops when other 
symmetry-allowed terms are added to $H_0({\bf k})$. In the following, we will show
that the transition from the 3D topological to trivial insulating states in class CI
(like the classes AIII, DIII, and CII) occurs through a gapless phase which is
generically a topologically protected nodal line semimetal.

The Hamiltonian in \cref{eq:H0CI} is not the most general Hamiltonian
in the class CI because many more terms can be  added to $H_0({\bf k})$
without breaking time-reversal and particle-hole symmetry. There are a seven types of couplings comprising nineteen 
 terms which preserve all the symmetries of class CI: $\eta^y 
\otimes {\bf b}, \eta^x \otimes {\bf c}, \eta^x \otimes {\bf d}$ and $\eta^x \otimes
{\bf b}$, 
$\eta^y \otimes {\bf c}, \eta^y \otimes {\bf d}$, ~$\delta$, where,
${\bf b} = (\sigma^z, \sigma^x, \tau^y \sigma^y)$, ${\bf c}=(-\tau^z\sigma^x,
\tau^z\sigma^z, \tau^x)$, ${\bf d} = (-\tau^x\sigma^x, \tau^x\sigma^z, -\tau^z)$, and 
$\delta=\eta^x\tau^0 \sigma^0$. We can further group these terms by looking at 
their transformation properties  under inversion, implemented by 
${\cal I}=\eta^y\tau^y$. The transformation properties of the various operators are listed 
in \cref{tab:table2}. The terms $(\eta^y \otimes {\bf b}, \eta^x \otimes {\bf c}, 
\eta^x \otimes {\bf d})$ are even under inversion, whereas the terms $(\eta^x \otimes {\bf b}, \eta^y
\otimes {\bf c}, \eta^y \otimes {\bf d}, \delta)$ are odd. Now we can 
write down the most general (minimal dimensional) Hamiltonian in class CI
\begin{align}\label{eq:HCI}
    H_{\text{CI}}({\bf k}) =& H_0({\bf k}) + \eta^y \otimes {\bf A}_b \cdot {\bf b} + \eta^x \otimes 
    {\bf A}_c\cdot {\bf c} \nonumber\\
    &+ \eta^x \otimes {\bf A}_d \cdot {\bf d} + \eta^x \otimes {\bf V}_b \cdot {\bf b} + \eta^y \otimes 
    {\bf V}_c \cdot {\bf c} \nonumber\\
    &+ \eta^y \otimes {\bf V}_d \cdot {\bf d} + V_{\delta} \delta ,
\end{align}
where $m$, $A$'s (inversion preserving), and $V$'s (inversion broken) can be any even
function of momenta. Without loss of generality, we can consider the following
quadratic dependence on momenta: $ X_{\alpha}({\bf k}) = X_{\alpha 0} - \sum_{ij} 
t^{(X_\alpha)}_{ij} k_i k_j$, where $X_{\alpha}$ is an element of the set $X = 
\{m, {\bf A}_b, {\bf A}_c, {\bf A}_d, {\bf V}_b, {\bf V}_c, {\bf V}_d,  V_{\delta} \}$, 
and all $X_{\alpha0}$, $t^{(X_\alpha)}_{ij}$ are reals. Since there are many terms, 
diagonalizing and finding the gapless points of $H_{\text{CI}}({\bf k})$ is 
a difficult task. However, in the following, we consider each of the perturbations 
individually and obtain enough information to establish the nature of the generic gapless phase of the Hamiltonian $H_{\text{CI}}({\bf k})$ unambiguously.

\begin{table}[ht]
  \begin{center}
    \begin{tabular}{ p{40mm}  p{20mm}  p{8mm} } 
      \hline 
      \hline
      \textbf{Operators} & \textbf{TR} & \textbf{I}\\ [1mm]
      \hline
      $\eta^x \otimes {\bf a}$ ~~ $(\Gamma_1, \Gamma_2, \Gamma_3) $ & $-1$ & $-1$ \\ [1mm]
      $\eta^x \otimes {\bf b} $ & $+1$ & $-1$ \\ [1mm]
      $\eta^x \otimes {\bf c} $ & $+1$ & $+1$ \\ [1mm]
      $\eta^x \otimes {\bf d}$ & $+1$ & $+1$ \\ [1mm]
      $\eta^x \otimes \tau^x \sigma^y$ & $-1$ & $+1$ \\ [1mm]
      $\eta^x \otimes \tau^z \sigma^y$ & $-1$ & $+1$ \\ [1mm]
      $\eta^x \otimes \tau^y \sigma^0$ & $-1$ & $-1$ \\ [1mm]
      $\eta^x \otimes \tau^0 \sigma^0$  & $+1$ & $-1$ \\ [3mm]
      
      $\eta^y \otimes {\bf a} $ & $-1$ & $+1$ \\ [1mm]
      $\eta^y \otimes {\bf b} $ & $+1$ & $+1$ \\ [1mm]
      $\eta^y \otimes {\bf c} $ & $+1$ & $-1$ \\ [1mm]
      $\eta^y \otimes {\bf d} $ & $+1$ & $-1$ \\ [1mm]
      $\eta^y \otimes \tau^x \sigma^y$ & $-1$ & $-1$ \\ [1mm]
      $\eta^y \otimes \tau^z \sigma^y$ & $-1$ & $-1$ \\ [1mm]
      $\eta^y \otimes \tau^y \sigma^0$ & $-1$ & $+1$ \\ [1mm]
      $\eta^y \otimes \tau^0 \sigma^0$ ~$(\Gamma_4)$  & $+1$ & $+1$ \\ [1mm]
      \hline 
    \end{tabular}
    \caption{Transformation properties of Dirac matrices (representation defined above \cref{eq:HCI}
     for class CI) which anticommute with ${\cal S} = \eta^z$.    \label{tab:table2}
}
  \end{center}
\end{table}

\subsection{\bf The perturbation $ \eta^y \otimes {\bf A}_{b} \cdot {\bf b}$:}
The Hamiltonian 
\begin{align}\label{eq:HAb}
H_{Ab}({\bf k}) = H_0({\bf k}) + \eta^y \otimes {\bf A}_{b} \cdot {\bf b}
\end{align}
can be readily diagonalized by squaring as we did before. We obtain the following  energy  spectrum 
\begin{align}\label{eq:EAb}
    E^2_{Ab}({\bf k}) = {\bf k}^2 + m^2  + {\bf A}_b^2 \pm  2\sqrt{ {\bf A}_b^2({\bf k}^2 + m^2) - 
    |{\bf A}_b \cdot {\bf k}|^2 }. 
\end{align}
Note that every band is doubly degenerate. Clearly, for gapless solutions, we must have 
\begin{align}
    {\bf k}^2 + m^2  + {\bf A}_b^2 =  2\sqrt{ {\bf A}_b^2({\bf k}^2 + m^2) - 
    |{\bf A}_b \cdot {\bf k}|^2 },
\end{align}
which can be rearranged to obtain 
\begin{align}
     \left({\bf k}^2 + m^2  - {\bf A}_b^2\right)^2  +  4|{\bf A}_b \cdot {\bf k}|^2 = 0.
\end{align}
Therefore for gapless solutions, ${\bf k}$, $M$ and ${\bf A}_b$ must satisfy the two 
conditions
\begin{align}
    {\bf k}^2 + m^2 = {\bf A}_b^2, ~~~~~ {\bf A}_b \cdot {\bf k} = 0.
\end{align}
The solution space, which is given by the intersection between the plane ${\bf A}_b
\cdot {\bf k} = 0$ and the surface $ {\bf k}^2 + m^2 = {\bf A}_b^2$ in 3D ${\bf k}$
space is generically a nodal line. As every band is doubly degenerate, the zero
energy nodal line is four-fold degenerate.

\subsection{\bf The perturbation $ \eta^x \otimes {\bf A}_{c} \cdot {\bf c}$:}
\noindent
The Hamiltonian 
\begin{align}\label{eq:HAc}
H_{Ac}({\bf k}) = H_0({\bf k}) + \eta^x \otimes {\bf A}_{c} \cdot {\bf c},
\end{align}
can be readily diagonalized to obtain 
\begin{align}\label{eq:EAc}
    E^2_{Ac}({\bf k}) = {\bf k}^2 + m^2  + {\bf A}_c^2 \pm  2|{\bf A}_c \cdot {\bf k}|. 
\end{align}
For  gapless solutions, we must have 
\begin{align}
     {\bf k}^2 + m^2  + {\bf A}_c^2 - 2|{\bf A}_c \cdot {\bf k}| = 0,  
\end{align}
which, after rearranging, reduces to the following
\begin{align}
    k_i^2 = A^2_{ci}, ~~~~ m({k_i^2 = A^2_{ci}})=0. 
\end{align}
For a generic $m({\bf k})=m_0-\sum_{ij}t_{ij}k_ik_j$, there will be no solutions to this pair of equations. 
 However, if one fine tunes $m_0$, one can obtain the solutions to be a pair of Dirac points located 
 at ${\bf k}_0= \pm {\bf A}_c$.
As in the case of CII, we will show that these Dirac points are not stable to generic 
perturbations. 

\subsection{\bf The perturbation $ \eta^x \otimes {\bf A}_{d} \cdot {\bf d}$:}
The Hamiltonian now is
\begin{align}
H_{Ad}({\bf k}) = H_0({\bf k}) + \eta^x \otimes {\bf A}_{d} \cdot {\bf d}. 
\end{align}
The energy spectrum and the gapless solution of $H_{Ad}({\bf k})$ are identical to the 
previous case but with the replacement ${\bf A}_{c}\to {\bf A}_{d}$. 

\subsection{\bf The perturbations $\eta^x \otimes {\bf V}_{b} \cdot {\bf b}$,  $\eta^y \otimes {\bf V}_{c} \cdot {\bf c}$,
 $\eta^y \otimes {\bf V}_{d} \cdot {\bf d}$:}

The individual perturbations $\eta^x \otimes {\bf V}_{b} \cdot {\bf b}$, 
$\eta^y \otimes {\bf V}_{c} \cdot {\bf c}$, $\eta^y \otimes {\bf V}_{d} \cdot {\bf d}$, 
when added to $H_0({\bf k})$ give identical results to the cases  above $ \eta^x
\otimes {\bf A}_{c} \cdot {\bf c}$,  $ \eta^y \otimes {\bf A}_{b} \cdot {\bf b}$
and $ \eta^y \otimes {\bf A}_{b}  \cdot {\bf b}$  but with the respective
replacements ${\bf A}_{c}\to {\bf V}_{b}$,  ${\bf A}_{b}\to {\bf V}_{c}$ and
${\bf A}_{b}\to {\bf V}_{d}$.

\subsection{\bf The perturbation $V_{\delta} \delta$:}
\noindent
The Hamiltonian 
\begin{align}
H_{V_{\delta}}({\bf k}) = H_0({\bf k}) + V_{\delta} \eta^x, 
\end{align}
can be easily diagonalized to obtain the following energy spectrum
\begin{align}
     E^2_{V_{\delta}}({\bf k}) = {\bf k}^2 + m^2  + {\bf V}^2_{\delta} \pm
     2V_{\delta}|{\bf k}|. 
\end{align}
Note that every band is doubly degenerate. For gapless solutions, we must have
\begin{align}
    {\bf k}^2 + m^2  +  V^2_{\delta} - 2V_{\delta}|{\bf k}|= 0, 
\end{align}
which, after rearrangement, reduces to 
\begin{align}
    {\bf k}^2 = V^2_{\delta}({\bf k}), ~~~~~ m^2({\bf k}) = 0. 
\end{align}
Clearly, the solution space, which is given by the intersection of two surfaces in
3D ${\bf k}$-space, is a nodal line for generic cases. Since every band is doubly degenerate,
the zero energy nodal loop is four-fold degenerate. Furthermore, the 1D AIII winding number associated with the four-fold degenerate
nodal loop is zero. 

Summarizing, when any single perturbation is added to
$H_0({\bf k})$, the eight-fold degenerate Dirac point goes to either a pair of four-fold
degenerate Dirac points or a four-fold degenerate nodal line with zero winding number. The double degeneracy of
every band, which leads to the four-fold degeneracy of the Dirac points and the nodal line,
can be traced back to the inversion symmetry of  the Hamiltonian. This is clear for the ${\bf A}$-type perturbations. For the ${\bf V}$-type perturbations, the Hamiltonian is invariant under the combined operation
of time-reversal $({\cal T})$ and inversion $({\cal I})$: ${\cal P} = {\cal I T}$,
which acts as ${\cal P} H({\bf k}){\cal P}^{-1} = H({\bf k})$ and satisfies ${\cal P}^2=-1$. Evidently, to remove all inversion-related symmetries, we will have to add an ${\bf A}$-type perturbation and a ${\bf V}$-type perturbation simultaneously. 

We now proceed to consider the generic Hamiltonian without any inversion-related symmetries via two paths: First we will consider a dominant ${\bf A}$-type perturbation which leads to a pair of four-fold degenerate Dirac points. We will then add a smaller ${\bf V}$-type perturbation and show that each Dirac point breaks up into a pair of two-fold degenerate nodal loops. In the second path, the dominant ${\bf A}$-type perturbation leads to a four-fold degenerate nodal loop. Upon adding a smaller ${\bf V}$-type perturbation, we find that this splits into a pair of two-fold degenerate nodal loops. In both cases, the two nodal loops have nontrivial, but opposite, winding numbers.

\subsection{\bf Four-fold degenerate Dirac point to two-fold degenerate nodal lines:}
We begin with the Hamiltonian $H_{Ac}({\bf k})$ in \cref{eq:HAc}
which describes four-fold degenerate Dirac points. Now we add $V_{\delta}\delta$ to $H_{Ac}({\bf k})$ to break the inversion symmetry (we have verified that adding any other $V$'s perturbation does not change the conclusion). Thus, 
we have the following Hamiltonian
\begin{equation}
\begin{aligned} \label{eq:HAcVd}
    H({\bf k}) = H_{Ac}({\bf k}) + V_{\delta}\delta. 
\end{aligned}
\end{equation}
Diagonalizing this Hamiltonian by the usual method of repeatedly squaring and separating terms, we find the following energy spectrum 
\begin{align}
      E^2({\bf k}) =& \left({\bf k}^2 + m^2  + {\bf A}^2_c + V^2_{\delta} \pm 2 {\bf k}\cdot {\bf A}_c \right) \nonumber\\
      &\pm 2 V_{\delta} \sqrt{{\bf k}^2  + {\bf A}^2_c  \pm 2 {\bf k}\cdot {\bf A}_c }.
\end{align}
Clearly, there are eight distinct eigenvalues. For gapless solutions, we must have 
\begin{align}
     \left({\bf k}^2 + m^2  + {\bf A}^2_c + V^2_{\delta} \pm 2 {\bf k}\cdot {\bf A}_c \right) = 2 V_{\delta} 
     \sqrt{{\bf k}^2  + {\bf A}^2_c  \pm 2 {\bf k}\cdot {\bf A}_c }.
\end{align}
This can be rearranged to obtain the following simple conditions for gapless solutions. 
\begin{align}
    \left({\bf k} \pm {\bf A}_c \right)^2 =V^2_{\delta}, ~~~~~ m^2({\bf k})=0. 
\end{align}
The solution space, given by the intersection of two surfaces in 3D ${\bf k}$-space,
is generically a nodal loop. Since there are generically eight distinct eigenvalues, the band crossing at the nodal loops is two-fold degenerate. Note that 
when $V_{\delta}\to 0$, the nodal loops approach Dirac points. The nodal loops always 
appear in pairs due to time reversal symmetry. We have computed their topological charge,
and we find that they carry opposite nonzero charges $W=\pm 1$. This is in contrast to the CII class, where the pair of nodal loops related by ${\cal T}$ carry the same winding number. We will elaborate on this difference shortly. 

\subsection{\bf Four-fold degenerate nodal line to two-fold degenerate nodal line:}
To demonstrate this path to the generic phase with two-fold degenerate nodal loops, we begin with the Hamiltonian $H_{Ab}({\bf k})$ in \cref{eq:HAb}
which describes a four-fold degenerate nodal line. Now we add the perturbation $V_{\delta}\delta$ to break inversion. The resulting  Hamiltonian is 
\begin{align} \label{eq:HAbVd}
    H({\bf k}) =  H_{Ab}({\bf k}) + V_{\delta}\delta. 
\end{align}
Diagonalizing this Hamiltonian, we find the following energy spectrum 
\begin{align}
      &E^2({\bf k}) = \left({\bf k}^2 + m^2  +  A^2_b + V^2_{\delta} \right) \nonumber\\
      &\pm 2 \sqrt{{\bf k}^2 V^2_{\delta}  + A^2_{b}({\bf k}^2 + m^2) -  A^2_{b} k_z^2 \pm 2 A_{b} k_z m V_{\delta}}.
\end{align}
For simplicity, we have rotated the vector ${\bf A}_b$ to align along the z-direction
${\bf A}_b=(0,0,A_b)$. Note that all eight eigenvalues are distinct, a reflection of
the fact that inversion symmetry is broken. For the gapless solution, we must have
\begin{widetext}
\begin{align}
    \left({\bf k}^2 + m^2  +  A^2_b + V^2_{\delta} \right) = 2\sqrt{{\bf k}^2 V^2_{\delta}  + A^2_{b}({\bf k}^2 + m^2) -  A^2_{b} k_z^2 \pm 2 A_{b} k_z m V_{\delta}},
\end{align}
\end{widetext}
which, after rearranging, reduces to 
\begin{align}
\left({\bf k}^2 + m^2 - A^2_b - V^2_{\delta}  \right)^2 + \left(A_b k_z \pm m V_{\delta} \right)^2 = 0.
\end{align}
Therefore for gapless solution ${\bf k}$, $m$, $A_b$ and $V_{\delta}$ must satisfy
the following two conditions 
\begin{align}
    {\bf k}^2 + m^2 = A^2_b + V^2_{\delta},  ~~~~ A_b k_z  = \pm m V_{\delta}
\end{align}
The solutions (the intersection of two surfaces) are  
generically two-fold degenerate nodal lines/loops related by ${\cal T}$.  We have 
computed their topological charge and find that they carry a nonzero opposite charge
$W =\pm 1$, as in  the previous case. 

From the above two cases, we find that the four-fold degenerate Dirac points, as well as the four-fold degenerate nodal lines, immediately go to a pair of two-fold generate nodal lines with a nonzero winding number when an inversion symmetry breaking perturbation is added. Since the two-fold degenerate nodal loops carry topologically nontrivial winding numbers, switching on  the remaining perturbations in the Hamiltonian $H_{\text{CI}}({\bf k})$ in \cref{eq:HCI} cannot gap out the nodal loops or turn the nodal loops into Dirac points immediately. Thus, we conclude that between the topological and trivial insulating states in class CI, there is a gapless phase that generically has pairs of  two-fold degenerate nodal loops related by ${\cal T}$.

\begin{figure}[ht]
\centering
\includegraphics[width=0.8\columnwidth]{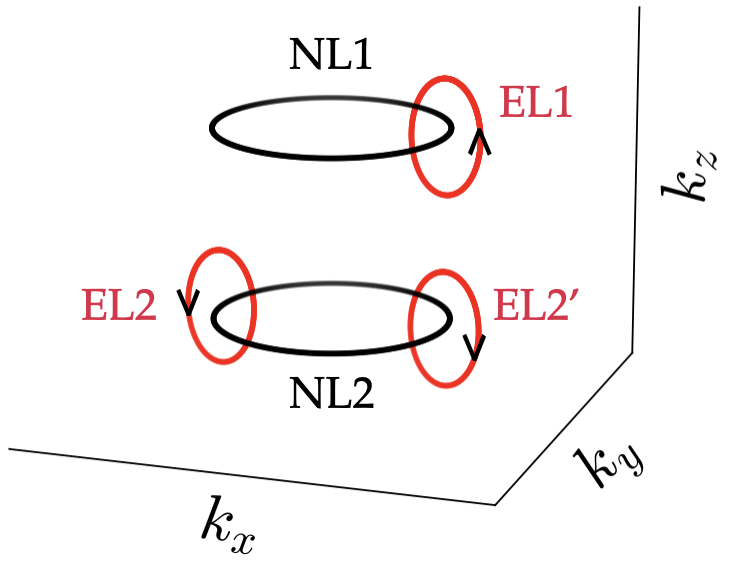}
\caption{A pair of nodal loops labeled NL1 and NL2. The enclosing loops are labeled EL1, EL2, and EL2$'$.   }
\label{fig:EL_Sense}
\end{figure}

\subsection{Topological invariant}

The previous classification \cite{Zhao_Wang_2013,Matsuura_2013,Zhao_Wang_2014} for the topological invariant for class CI in gapless phases is $\mathbb{Z}$, and the invariant is the chiral winding number \cref{eq:invariant}. Thus, the only difference between our approach and the previous one for class CI is that our loops are unrestricted, and do not have to be time reversal symmetric.

Let us now understand the difference between classes CI and CII, where the nodal loops related by ${\cal T}$ have either opposite or identical winding numbers, respectively. The very existence of the winding number is a direct consequence of chiral symmetry, implemented by ${\cal S}=\eta^z$ in both classes. The key is to look at how ${\cal S}$ transforms under ${\cal T}$. In class CII, ${\cal T}{\cal S}{\cal T}^{-1}={\cal S}$. As we will see very shortly, this implies that the two nodal loops related by ${\cal T}$ have the same winding number. For class CI we have ${\cal T}{\cal S}{\cal T}^{-1}={-\cal S}$, implying opposite winding numbers for the two nodal loops related by ${\cal T}$. The same will be true in class DIII, for the same reason.

In order to proceed we need to define the winding numbers of the two loops (NL1 and NL2) related by ${\cal T}$ for a generic configuration.  It is best to start with a simple case in which NL1 and NL2 are in the $k_xk_y$ plane, and separated by $k_z$, as shown in Fig. \ref{fig:EL_Sense}.  Take one of the pair of nodal loops (NL1, say) and define an enclosing loop (called EL1, say) winding around NL1 only, in a certain sense. Applying ${\cal T}$ to the enclosing loop we obtain another enclosing loop that winds around NL2. The sense of the enclosing loop EL1 defines a definite sense for the second enclosing loop (EL2) under ${\cal T}$. Now we ``slide" the loop EL2 around NL2 so that it lies directly below EL1. This defines a new enslosing loop EL2$'$. One can see that the sense in which EL1 and EL2 are traversed is opposite. We define the winding number of NL2 as the {\it opposite} of the result obtained from EL2$'$. This definition turns out to naturally correspond to the degeneracies of zero-energy surface states. These are the two winding numbers which are either the same (class CII) or opposite (classes CI and DIII).

Now we come to the computation of the winding number, \cref{eq:invariant}, repeated here for convenience
\begin{align}\label{eq:invariant_repeat}
    W = \frac{1}{2\pi i} \int_{0}^{2\pi} dk ~ \textrm{Tr}\left(Q^{-1} \partial_k Q\right).
\end{align}
Recall the way the matrix $Q$ is defined: One goes to a basis where $\eta^z$ is ${\cal S}$, implying that the 
Hamiltonian can be expressed as 
\begin{align}
    \begin{pmatrix}
        0&Q\\
        Q^{\dagger}&0\\
    \end{pmatrix}
\end{align}
Now we are ready to find the relation between the winding numbers of NL1 and NL2. Under ${\cal T}$ there is the sense of the loop changes sign as we saw in the previous paragraph. Also, ${\cal T}$ comes with complex conjugation, which produces another sign reversal as per \cref{eq:invariant_repeat}. In class CII, there are no further sign reversals, and thus the winding numbers of NL1 and NL2, as defined above, are identical. In classes CI and DIII, ${\cal T}$ comes with an additional $\eta^x$. This changes $Q\rightarrow Q^{\dagger}$, which is yet another complex conjugation, implying one more sign reversal. This is why the winding numbers of NL1 and NL2 are opposite in classes CI and DIII. 

The senses of the winding numbers were defined above for a particularly simple pair of ${\cal T}$-related nodal loops. Given a generic pair of ${\cal T}$-related nodal loops we proceed as follows: First we find the straight line going through both nodal loops such that the overlap of the projections of the two nodal loops on to the surface perpendicular to the axis is maximized. We then proceed as described in the paragraphs above. 

Let us note that the relation between the winding numbers is controlled by $\epsilon_S=\pm1$, defined by ${\cal T}{\cal S}{\cal T}^{-1}=\epsilon_S {\cal S}$. $\epsilon_S$ is dependent on $\epsilon_T$ and $\epsilon_C$, defined by  ${\cal T}^2=\epsilon_T$ and ${\cal C}^2=\epsilon_C$. Since we have chosen to implement ${\cal S}={\cal T}{\cal C}$ by $\eta^z$, we know that ${\cal S}^2=1$. Thus,
\begin{align}
  & {\cal S}^2={\cal T}{\cal C}{\cal T}{\cal C}=1 \nonumber\\
   \Rightarrow ~ & {\cal T}{\cal S}^2=\epsilon_T {\cal C}{\cal T}{\cal C}={{\cal T}}\nonumber\\
   \Rightarrow ~ & {\cal C}{\cal T}{\cal S}^2={\cal C}{\cal T}=\epsilon_C\epsilon_T{\cal T}{\cal C}.
   \label{eq:TCComm}
\end{align}
This is the ``commutation relation" of ${\cal T}$ and ${\cal C}$. Now, let ${\cal T}{\cal S}{\cal T}^{-1}=\epsilon_S{\cal S}$, where $\epsilon_S=\pm1$. Since ${\cal T}^{-1}=\epsilon_T {\cal T}$, this implies
\begin{align}
{\cal T}{\cal S}{\cal T}^{-1}=
    \epsilon_T{\cal T}{\cal S}{\cal T}=\epsilon_T{\cal T}{\cal T}{\cal C}{\cal T}={\cal C}{\cal T}=\epsilon_S{\cal T}{\cal C}
\end{align}
Comparing with \cref{eq:TCComm} we finally obtain the desired relation.
\begin{align}
    \epsilon_S=\epsilon_T\epsilon_C
    \label{eq:esetec}
\end{align}
For class CII, $\epsilon_T=\epsilon_C$, which implies that chirality is even under ${\cal T}$, and thus the two nodal loops related by ${\cal T}$ have the same winding number. However, for classes CI and DIII, $\epsilon_T=-\epsilon_C$, which means that chirality is odd under ${\cal T}$, implying that the two nodal loops related by ${\cal T}$ have opposite winding numbers.

Despite the fact that both our classification and the previous one \cite{Zhao_Wang_2013,Matsuura_2013, Zhao_Wang_2014} use the winding number in class CI, the fact that our loops are not constrained to be centrosymmetric gives our approach a better ability to distinguish the presence/absence of gapless drumhead modes at specific points on the surface BZ when open surfaces are present. 
We will elaborate on this in \cref{Sec:Surface}.

\vspace{0.2cm}

\begin{table}[ht]
  \begin{center}
    \begin{tabular}{p{40mm} p{20mm} p{8mm}} 
      \hline 
      \hline
      \textbf{Operators} & \textbf{TR} & \textbf{I}\\ [1mm]
      \hline
      $\eta^x \otimes {\bf \boldsymbol{\sigma}}$ ~~ $(\gamma_1, \gamma_2, \gamma_3) $ & $-1$ & $+1$ \\ [1mm]
      $\eta^x \otimes \sigma^0 $ & $+1$ & $-1$ \\ [1mm]
      $\eta^y \otimes {\bf \boldsymbol{\sigma}}$  & $-1$ & $+1$ \\ [1mm]
      $\eta^y \otimes \sigma^0 $ ~ $(\gamma_4)$ & $+1$ & $+1$ \\ [1mm]
      \hline 
    \end{tabular}
    \caption{Transformation properties of Dirac matrices (representation given in \cref{eq:H0DIII} 
     for the class DIII)
    which anticommute with ${\cal S} = \eta^z$.    \label{tab:table3}
}
  \end{center}
\end{table}

\section{Class DIII}
\label{Sec:DIII}
\noindent
Hamiltonians in class DIII have 
${\cal T}^2=-1, {\cal C}^2=1$. A minimal Hamiltonian which describes the transition
between 3D topological and trivial insulating states and also satisfies the above
symmetries is a four-component Dirac Hamiltonian
\begin{align}\label{eq:H0DIII}
    H_0({\bf k}) = \eta^x ~{\bf k}.{\boldsymbol \sigma} + m\eta^y.
\end{align}
Here $\sigma$’s and $\eta$’s act on spin and sublattice space, respectively. The 
time-reversal and particle-hole operations are realized by ${\cal T}=\eta^x \sigma^y K$ 
and ${\cal C}= i\eta^y \sigma^y K$ respectively. They clearly satisfy ${\cal T}^2=-1, 
{\cal C}^2=1$, confirming that $H_0({\bf k})$ belongs to class DIII.
The chiral symmetry operation is simply given by the product of particle-hole
and time-reversal ${\cal S} = {\cal T}{\cal C} = \eta^z$. The Hamiltonian is also 
symmetric under the inversion ${\cal I}=\eta^y$. The transition from topological
to trivial insulating phase occurs through a Dirac point at the origin ${\bf k}=0$.
In the following, we will see that this Dirac point is not stable. It will immediately go
to nodal loops when other symmetry-allowed terms are added to $H_0({\bf k})$. 
As usual, the Hamiltonian in \cref{eq:H0DIII} is not the most general class
DIII Hamiltonian, because  there is one more symmetry-allowed term which can be added to $H_0({\bf k})$, namely  
$\eta^x\sigma^0$. Therefore, the most general $4\times4$ Hamiltonian in class DIII  is 
\begin{align}\label{eq:HDIII}
 H_{\text{DIII}}({\bf k}) = \eta^x ~{\bf k}.{\boldsymbol \sigma} + m \eta^y + V \eta^x. 
\end{align}
This Hamiltonian can be easily diagonalized to obtain the following energy spectrum. 
\begin{align}
    E^2({\bf k}) = {\bf k}^2 + m^2 + V^2 \pm 2V |{\bf k}|. 
\end{align}
The two parameters $m$ and $V$ can be any even function of the components of ${\bf k}$; a minimal choice
would be $m\to m({\bf k})= m_0 - \sum_{ij} t^{(m)}_{ij} k_i k_j$, ~ $V\to V({\bf k})=
V_0 - \sum_{ij} t^{(V)}_{ij} k_i k_j$, where $m_0, V_0,  t^{(m)}_{ij}$ and $t^{(V)}_{ij}$
are all real. For the energy spectrum to be gapless, ${\bf k}$, $m({\bf k})$ and $V({\bf 
k})$ must satisfy the following conditions 
\begin{align}
m({\bf k}) = 0, ~~~~~~~ {\bf k}^2 = V^2({\bf k}).
\end{align}
The solution space, which is given by the intersection of two surfaces in the 3D
${\bf k}$ space is generically a nodal line. The nodal loops always appear in pairs
due to time reversal symmetry. As usual, small deformations of the Hamiltonian cannot gap out the nodal loops.  To see whether the loops enjoy topological 
protection, we compute the AIII winding number using  \cref{eq:invariant}. We find that the pair
of nodal  loops related by ${\cal T}$ carry opposite topological charge $W=\pm 1$, in accordance with \cref{eq:esetec}
Therefore, the transition between 3D topological and trivial insulating states in class 
DIII happens through a gapless phase which is generically a nodal line semimetal. Since
the nodal loops carry a nontrivial winding number, they are topologically protected and 
stable against gap opening. 

As in the case of class CI, the previous classification in DIII is 
$\mathbb{Z}$ \cite{Zhao_Wang_2013,Matsuura_2013, Zhao_Wang_2014},  with the invariant being the winding number \cref{eq:invariant}. Our topological charge is also the winding number, with the difference that our loops are not restricted to be centrosymmetric. Surface states 
in class DIII which either overlap completely or do not overlap at all were discussed in Ref. \onlinecite{Matsuura_2013}. Our approach, presented next in  \cref{Sec:Surface}, allows us to directly use our bulk invariant to predict the properties of surface states.

\begin{figure*}
     \centering
     \includegraphics[width=0.95\linewidth]{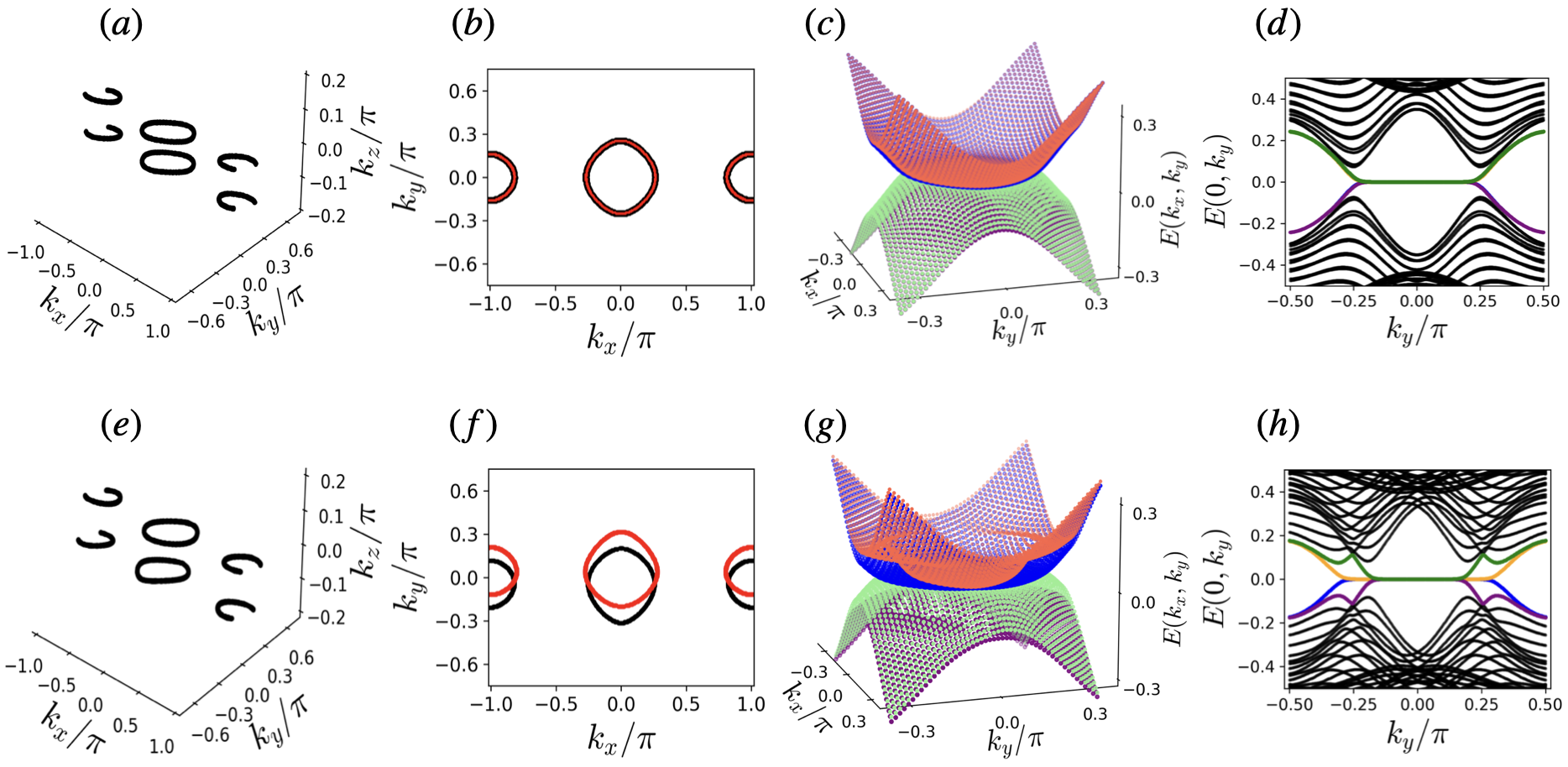}
     \caption{Bulk nodal loops, their surface projections (on $k_x$-$k_y$ BZ), and the 
     associated surface states of the lattice models in class CII. For the  first row,
     the parameters $m=0.1$, $t_1=-0.1$, $t_2=-0.2$, $t_3=1.1$, ${\bf A}_a = (0,0.1, 
     1.0)$, $A_5=0.2$, $V_{\epsilon}=0.2$, and all other terms are kept zero, are such 
     that the projections of the pairs of nodal loops (in $(a)$), related by
     time-reversal symmetry, almost overlap on $k_x$-$k_y$ surface BZ (in $(b)$).
     Figure $(c)$, which presents the low energy spectrum of the system taken finite
     along the $z$-direction, shows the flat drumhead surface states which are associated
     with  the bulk nodal loops around the origin $k_x=k_y=0$. $(d)$ Shows surface states along the 
     cut $k_x=0$. In the second row, the
     parameters are the same as the first row but with the following nonzero $V$ terms:
     ${\bf V}_p=(0.1,0,0)$, ${\bf V}_b =(0,0.1,0)$, ${\bf V}_{b'} =(0,0.15,0)$.
     Now projections of the pair of nodal loops (in $(e)$) do not overlap completely on the
     $k_x$-$k_y$ surface BZ (in $(f)$). Figure $(g)$, which presents the low energy spectrum
     of the system taken finite along the $z$-direction, shows the flat drumhead 
     surface states, which are four/two-fold degenerate (clearly shown in Fig. (h)) in the
     overlapping/non-overlapping regions in the surface BZ, respectively. Figure (h) shows low energy 
     spectrum along the cut $k_x=0$. }
\label{fig:CII_Surface}
\end{figure*}

\section{Surface States}
\label{Sec:Surface}

The salient experimental signature of TNLSMs is the existence of  topologically protected gapless drumhead surface states in a finite region of the surface BZ \cite{Hosen_2020,Bian2016n,wang2017,Lou_2018,Zhou2019,Belopolski_2019,Lv2021,Chen2021,Stuart2022,Gao2023}. 
In our classification, the nodal loops in classes AIII, CII, CI, and DIII are characterized 
by the winding number on a loop that ``winds around" the given nodal loop. Recall that this winding number 
is an invariant belonging to class AIII, because the Hamiltonian restricted to an arbitrary linking loop 
only has chiral symmetry, and thus belongs to AIII. 
Thus, by the bulk boundary correspondence in class AIII, there must be gapless surface states 
associated with nodal loops in AIII,
CII, CI, and DIII. Since there exist multiple k-values through which an enclosing 
loop can be drawn, there must be a finite region of gapless surface states in
surface BZ. For slabs of large $z$-thickness in lattice units, the drumhead states are very close to exactly degenerate, and consist of a symmetric and an antisymmetric combination of zero-energy states localized on the top and bottom surfaces.

To get a picture of the surface states associated with the nodal loops, we will 
consider specific lattice models for each of the above chiral classes. The surface states
associated with the nodal loops in class AIII are discussed in the companion paper Ref. \cite{MainPaper},
so here we focus on the remaining three classes, which are more interesting due to the
additional time reversal and particle-hole symmetries.

One can modify the continuum Hamiltonian to obtain a lattice model by replacing linear terms
in $k_i$ by $\sin{k}_i$ and the quadratic terms  in  $k$ by a linear combination of $\cos{k}_i$ ($M({\bf k})=m-\sum_it_i\cos{k}_i$). Following this
prescription, we obtain a model defined on a
cubic lattice
\begin{align}\label{Eq:LatticeModel}
    H = \sum_{{\bf n}, j} c^{\dagger}({\bf n})~ 2 {\bf Q} ~ c({\bf n}) - \left(c^{\dagger}({\bf n} +  \hat{e}_j) 
    ~ {\bf T}_j ~ c({\bf n}) + H.c \right), 
\end{align}
where $n = (n_1,n_2,n_3)$  ($n_i$ integer), denote the lattice sites, $\hat{e}_j$
is the unit vector along $j^{th}$ direction, and the lattice constant has been set to unity.
The matrix dimension $N$ of the onsite ${\bf Q}$ and hopping terms ${\bf T}$
is identical to the matrix dimension of the Dirac 
Hamiltonian in the given symmetry class. The $c^{\dagger}({\bf n})$ and $c({\bf n})$ are
the $N$-component Dirac fermion creation and annihilation operators. Note
that the onsite term ${\bf Q}$ includes all the momentum-independent terms allowed by
the symmetry class. The hopping matrices ${\bf T}_j$, $j=1, 2, 3$, may be expressed in
a generic form ${\bf T}_j = t_j  \Gamma_4 + i\Gamma_j $, where $i$ is the imaginary 
unit and $t_j$'s are real numbers. The $\Gamma_j $ and $\Gamma_4$ are the gamma
matrices satisfying the Clifford algebra $\{\Gamma_i, \Gamma_j \} = 2 \delta_{ij}$, $i,
j=1, 2, 3, 4$. In the following, we will explicitly specify $\Gamma_j, {\bf Q}$ and 
${\bf T}_j$ for each of the three classes CII, CI, and DIII to obtain the lattice models,
which we will then use to compute the surface states associated with the bulk nodal loops. 

\begin{figure*}
     \includegraphics[width=0.99\linewidth]{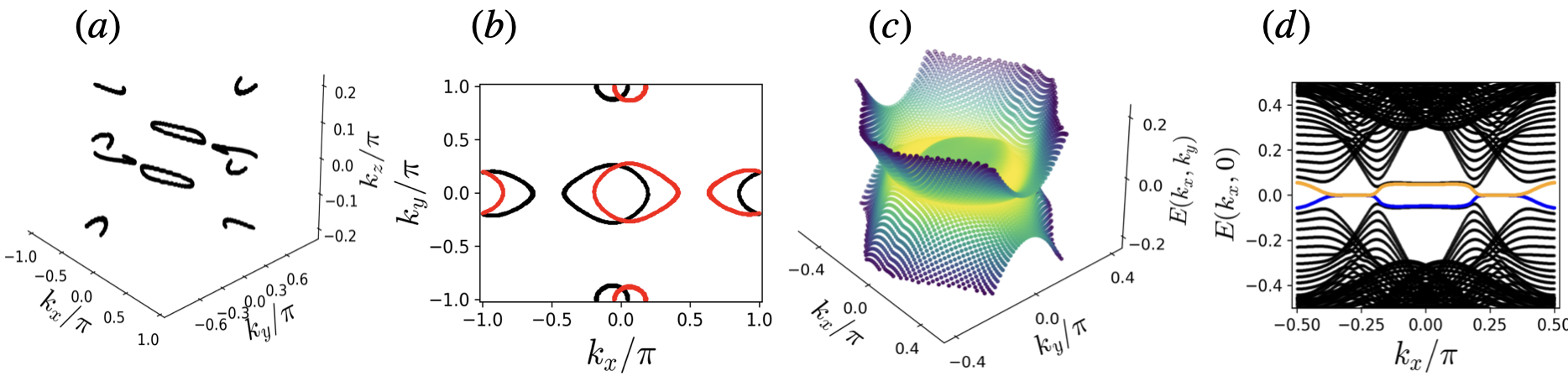}
     \caption{ $(a)$ Bulk nodal loops and $(b)$ their surface projections (on $k_x$-$k_y$
     BZ), of the lattice model in class  CI. The parameters are $m=0.1$, $t_1=-0.1$, 
     $t_2=-0.2$, $t_3=0.8$,${\bf A}_b = (0,0.1, 0.8)$, $V_{\delta}=0.4$, ${\bf V}_b=
     (0.2,0,0)$, ${\bf V}_c=(0,0.1,0)$, and the remaining parameters are set to zero.
     Figures $(c)$ and $(d)$ show the surface states associated with the  bulk nodal loops
     around $k_x =k_y=0$. Figure $(c)$ presents the low energy spectrum of the slab. Note that he flat  zero-energy 
     drumhead surface states exist only in the non-overlapping region. There
     are no zero energy surface states in the overlapping region. For clear
     visualization, a section of the surface states (highlighted in color) for fixed $k_y=0$ are 
     separately shown in  $(d)$.}
\label{fig:CI_Surface}
\end{figure*}

\subsection{Class CII}

In class CII, the minimal Dirac Hamiltonian is $8\times8$, therefore
$c^{\dagger}({\bf n})$ and $c({\bf n})$ are 8-component  Dirac fermion
creation and annihilation operators. From the Dirac Hamiltonian in 
\cref{eq:HCII}, we have the following $\Gamma$ matrices 
\begin{align*}
  \Gamma_1 = \eta^x \tau^z \sigma^x,~  \Gamma_2 = \eta^x \tau^z \sigma^y,~  \Gamma_3 = \eta^x \tau^y,~  
  \Gamma_4 = \eta^x \tau^x. 
\end{align*}
Recall that the Pauli matrices $\eta^a$, $\tau^a$ and $\sigma^a$ act on the
sublattice, orbital, and  spin spaces respectively. The momentum independent
matrix ${\bf Q}$ is given by 
\begin{align*}
    {\bf Q}  &=  m \eta^x \tau^x + \mu \eta^x \otimes \gamma^0 + \eta^y\otimes{\bf A}_a\cdot{\bf a} 
          +A_5\eta^y\otimes\gamma^5 \\ & + 
     \eta^y \otimes {\bf V}_{b} \cdot {\bf b}  +  \eta^y \otimes 
     {\bf V}_{b'} \cdot {\bf b}'  + \eta^x \otimes {\bf V}_{p} \cdot {\bf p} +  
      V_{\epsilon} \eta^x \otimes \epsilon,
\end{align*}
where all the terms on the right-hand side are defined in \cref{Sec:CII}. 

To illustrate the various possible geometries of the surface states associated with bulk topological nodal 
loops, we take two representative sets of parameters (i) $m=0.1$, $t_1=-0.1$, 
$t_2=-0.2$, $t_3=1.1$ (recall $M({\bf k})=m-\sum_it_i\cos{k}_i$), ${\bf A}_a = (0,0.1, 1.0)$,  $A_5=0.2$, $V_{\epsilon}=0.2$
and the remaining terms are zero;  (ii) the same as (i) but with the following additional 
terms ${\bf V}_p =(0.1,0,0)$, ${\bf V}_b =(0,0.1,0)$, ${\bf V}_{b'} =(0,0.15,0)$.
For both sets of parameters, the lattice model describes a gapless
nodal line semimetal; the nodal loops  around  $k_z=0$ are shown in 
\cref{fig:CII_Surface}. We will take the open surface to be the $xy$-plane. The lattice model is solved on a slab that is finite in the $z$-direction, and has periodic boundary conditions in the $x$ and $y$ directions, ensuring that $k_x,k_y$, and thus the surface BZ, are well-defined. For the first set, the pair of nodal loops related
by ${\cal T}$ have projections on to the surface BZ that overlap almost perfectly. The   low energy spectrum as a function of 
$k_x$ and $k_y$  is shown in \cref{fig:CII_Surface}(c). There are four-fold degenerate zero-energy surface
states in the entire region that is bounded by the projection of bulk nodal loops in
the $k_x$-$k_y$ surface BZ.

To see a slightly different geometry, we turn to the second set of parameters. Here there is a region of the surface BZ where the projections of the two nodal loops
related by ${\cal T}$ do not overlap. Now it is clear that the drumhead states are four-fold degenerate in the region of overlap, but only two-fold degenerate in the non-overlapping region.  The bulk nodal loops and their surface states for the second case are 
depicted in the bottom panel of \cref{fig:CII_Surface}.

We can easily understand these degeneracies as follows: By virtue of having a nonzero winding number, each nodal loop gives rise to a two-fold degenerate drumhead state, with the two zero-energy states residing on the top and bottom of the slab. In CII, the two nodal loops related by ${\cal T}$ have the same winding number from \cref{eq:esetec}, and hence in the region of overlap on the surface BZ, the degeneracy is increased by a factor of 2. As we will see shortly, in classes CI and DIII, the logic of \cref{eq:esetec} implies that there are no drumhead states in the region of overlap of two nodal loops related by $\mathcal{T}$. 

\begin{figure*}
     \centering
     \includegraphics[width=0.99\linewidth]{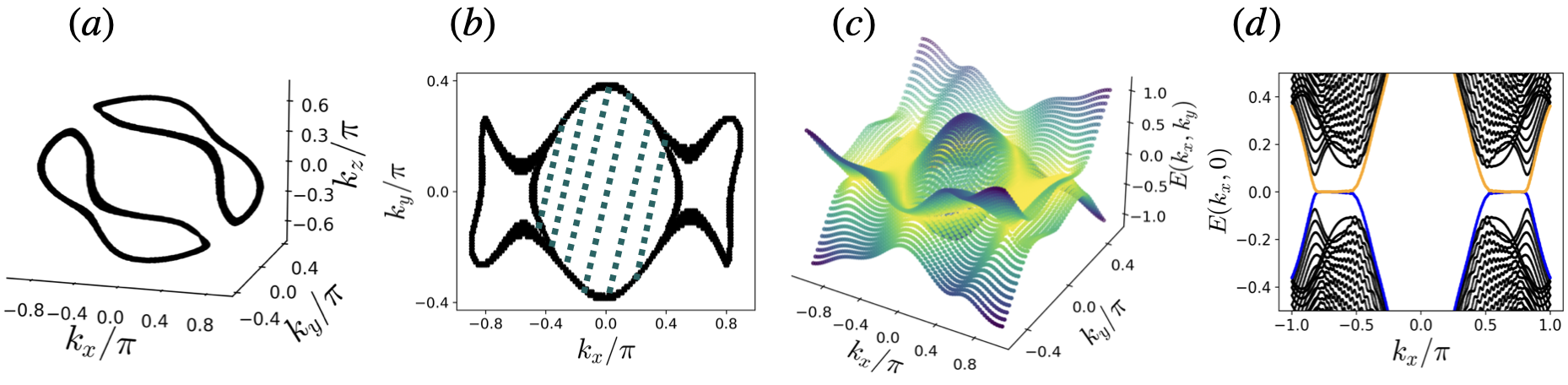}
     \caption{$(a)$ Bulk nodal loops and $(b)$ their surface projections on the $k_x$-$k_y$ 
     surface BZ, of the lattice model in class DIII (Eq. \ref{eq:DIII_Latice}). Their projections 
     of the $k_x$-$k_y$ surface BZ overlap in a finite region (shaded region in $(b)$). 
     Figures $(c)$ and $(d)$ shows the surface states associated with the  bulk nodal loops.
     Figures $(c)$  presents the low energy  spectrum of the system taken 
     finite along the $z$-direction. The zero energy surface states associated 
     with  the bulk nodal loops exist only in the nonoverlapping region of projection on the 
     $k_x$-$k_y$ surface BZ. For clear visualization, a section of the surface states (highlighted 
      in color) for fixed ky = 0 is separately shown in $(d)$. The surface spectrum is gapped 
      in the overlapping region but it is gapless  and flat in the nonoverlapping regions of 
      projections of the nodal loops. }
     \label{fig:DIII_Surface}
\end{figure*}

\subsection{Class CI}

As in class CII, the minimal Dirac Hamiltonian in class CI is $8\times8$.
From \cref{eq:HCI}, we read off the four  $\Gamma$ matrices 
\begin{align*}
  \Gamma_1 = \eta^x \tau^y \sigma^z,~  \Gamma_2 = \eta^x \tau^y \sigma^x,~  \Gamma_3 = \eta^x \sigma^y,~  
  \Gamma_4 = \eta^z. 
\end{align*}
The momentum independent
matrix ${\bf Q}$ for this class is given by  
\begin{align*}
{\bf Q} = & ~  m\eta^z + \eta^z \otimes {\bf A}_b \cdot {\bf b} + \eta^x \otimes {\bf A}_c\cdot {\bf c} + 
             \eta^x \otimes {\bf A}_d \cdot {\bf d} \\ & + \eta^x \otimes {\bf V}_b \cdot {\bf b} + 
    \eta^z \otimes {\bf V}_c \cdot {\bf c}  + \eta^z \otimes {\bf V}_d \cdot {\bf d} + V_{\delta} \delta. 
\end{align*}
For the definition of the various  terms on the right-hand side, see \cref{Sec:CI}. 

Let us consider a representative choice  of parameters  $m=0.1$, $(t_1, t_2, t_3)=
(-0.1, -0.2, 0.8)$ (recall $M({\bf k})=m-\sum_it_i\cos{k}_i$), ${\bf A}_b = (0,0.1, 0.8)$, $V_{\delta}=0.4$, ${\bf V}_b=(0.2,0,0)$, 
${\bf V}_c=(0,0.1,0)$, with the remaining parameters being set to zero. The lattice model has pairs of nodal loops around
$k_z=0$, as shown in \cref{fig:CI_Surface}(a).  Their surface projections on the $k_x$-$k_y$ surface BZ
are shown in \cref{fig:CI_Surface}(b).  The low energy
spectrum as a function of  $k_x$ and $k_y$ for the slab geometry is shown  in \cref{fig:CI_Surface}(c)-(d). We find that
there are no gapless surface states in the region of the surface BZ where the projections of the nodal loops overlap. Clearly, this is because they have opposite winding numbers, in accordance with \cref{eq:esetec}. However, there are two-fold degenerate drumhead states in the
regions of the surface BZ where the nodal loop projections do not overlap.

\subsection{Class DIII}

The lattice model for class DIII  in Eq. \ref{Eq:LatticeModel} is not flexible enough to illustrate the full range of possibilities for drumhead 
states because it has $C_2$ symmetry around each axis, which forces the projections of the nodal loops on the surface BZ
will always overlap. Surface states are better illustrated when the projections of the 
nodal loops have both overlapping and nonoverlapping regions on the surface BZ.

To make the model generic, we break this two-fold rotation symmetry by adding next nearest neighbour hoppings. 
\begin{widetext}
\begin{align}\label{eq:DIII_Latice}
    H({\bf k}) = \sum_i \sin{k_i} \Gamma_i + \left(m - \sum_i t_{mi} \cos{k_i} - \sum_{ij, i\neq j} 
    t_{ij}\cos{(k_i + k_j)} + t^{'}_{ij}\cos{(k_i - k_j)} \right)\Gamma_4 
    + \left(V - \sum_i t_{vi} \cos{k_i}\right) \eta^x \sigma^0
\end{align}
\end{widetext}
where $\Gamma_1 = \eta^x \sigma^x, ~ \Gamma_2 = \eta^x \sigma^y,  ~\Gamma_3 = \eta^x \sigma^z$, 
$\Gamma_4 = \eta^y$, and $t_{ij}=t_{ji}$, $t^{'}_{ij}=t^{'}_{ji}$. The Pauli matrices $\eta$'s, 
and $\sigma$'s act on the sublattice and spin spaces, respectively. The choice $t_{mi}\neq 
t_{mj}, t_{vi}\neq t_{vj}$, and  $t_{ij} \neq t^{'}_{ij}$ make sure that crystalline symmetries 
are broken. 

Let us consider a representative choice  of parameters $m=0.4, V=0.6$, ${\bf t}_{m}=(0.8, 0.5, 1.0)$, 
${\bf t}_v = (-0.5, -0.8, -0.2)$ and $t_{xy}=0, t^{'}_{xy}=0$, $t_{xz}=0.6, t^{'}_{xz}=0.01$, 
$t_{yz}=0.3, t^{'}_{yz}=0.01$. The lattice model has pairs of nodal loops  as shown in 
\cref{fig:DIII_Surface}(a). Their surface projections on the $k_x$-$k_y$ surface BZ
are shown in \cref{fig:DIII_Surface}(b). The low energy spectrum as a function of  $k_x$ and $k_y$ 
for the slab geometry is shown  in \cref{fig:DIII_Surface}(c)-(d). Like the surface states in class 
CI, we find that there are no gapless surface states in the region of the surface BZ where the 
projections of the nodal loops overlap. Clearly, this is because they have opposite winding numbers, 
in accordance with \cref{eq:esetec}. However, there are two-fold degenerate drumhead states in the
regions of the surface BZ where the nodal loop projections do not overlap.

\section{Summary \& outlook}
\label{Sec:Summary}

The fact that gapless states can have topological protection has been known for more than a decade \cite{Murakami2007,Wan2011,Burkov_Balents_2011,Burkov2011}. While initial work focused on Weyl and Dirac semimetals, there has been quite a bit of work on nodal loop semimetals as well \cite{Burkov2011,Fang2016,Chang2017,Heikkila2015,das2020,das2022}. In any classification scheme the ``internal" symmetries (time reversal ${\cal T}$, charge conjugation ${\cal C}$, and their product, the chirality ${\cal S}={\cal T}{\cal C}$) are crucial \cite{Zhao_Wang_2013,Matsuura_2013,Zhao_Wang_2014}. There can be additional ``lattice" symmetries beyond translations as well \cite{Chiu2014, Chen2015TopologicalCM,Kim2015SurfaceSO, Yamakage2016, Bian2016, Bian2016n, Sun2017TopologicalNL,  Nie2019, Wang2021TopologicalNC,Fang2015,Kim2015, Weng2015TopologicalNS, Yu2015TopologicalNS, Huang2016,Fang2015,Schoop2016,Hong2018MeasurementOT,Meng2020ANH, Wang2021SpectroscopicEF,Xie2021KramersNL}, such as mirror symmetries, non-symmorphic symmetries, or $SU(2)$ spin-rotation symmetry. In a semimetal, a powerful way to see if the gapless state is topologically protected is to enclose the Fermi points/lines in a lower dimensional subspace of the Brillouin zone, and see if the gapped Hamiltonian on that enclosing subspace is topologically nontrivial.  

In this work, we have focused on the three-dimensional case when no lattice symmetries other than translations are present, and when ${\cal S}$ is a symmetry of the Hamiltonian. A previous classification \cite{Zhao_Wang_2013,Matsuura_2013,Zhao_Wang_2014} for this case uses enclosing surfaces which have all the symmetries of the Hamiltonian. Specifically, if the Hamiltonian in question is symmetric under ${\cal T}$, the enclosing surface must have both ${\bf k}$ and $-{\bf k}$, that is, it must be centrosymmetric. 

The primary difference between our classification and the previous one is that our enclosing surfaces are generic, and need not be centrosymmetric. Our first result is  that if there is a topologically nontrivial insulator in a chiral class in three dimensions, then there is generically a semimetal phase between the topologically nontrivial and the topologically trivial insulating phases. This applies to classes AIII, CI, CII, and DIII. Class BDI also enjoys chiral symmetry, but it does not have a topologically nontrivial insulator in three dimensions. Note that the implication goes only one way: BDI may have topologically protected nodal loop semimetal phases, but we are not able to guarantee it by our logic. We are further able to show by contradiction that the semimetal phase in a chiral class cannot have isolated, topologically protected, point nodes. If there is such a point node, we enclose it in an arbitrary two-dimensional surface. The Hamiltonian restricted to this surface has only chiral symmetry, and thus belongs to class AIII. However, class AIII has a trivial topology in two dimensions, which shows the contradiction. Our second result is the natural sequel to the first. We show that the generic semimetal must be a nodal loop semimetal. Now the enclosing ``surface" is an arbitrary loop in the Brillouin zone that winds around the putative nodal loop. The Hamiltonian restricted to the enclosing loop has only ${\cal S}$, and is in class AIII. Class AIII does have nontrivial topology in one dimension labeled by the winding number, implying that if the nodal loops have nonzero winding number they are topologically protected. It is worth noting that we use the winding number to classify semimetals in all the four chiral classes we consider (AIII, CI, CII, DIII), making it a universal tool for investigating nodal loop phases in Hamiltonians with chiral symmetry. It turns out that in class AIII (no ${\cal T}$ or ${\cal C}$, only ${\cal S}$) single nodal loops can appear, but in the other three classes, time reversal symmetry forces generic nodal loops to appear in pairs. We have shown that if ${\cal T}^2={\cal C}^2$, as happens in class CII, the two loops related by ${\cal T}$ have the same winding number. On the other hand, if ${\cal T}^2=-{\cal C}^2$, which occurs in classes CI and DIII, the two loops have opposite winding numbers. We substantiate all these theoretical results by explicit calculations in minimal models in each symmetry class. 

For classes AIII, CI, and DIII, both our approach and the previous one give essentially the same classification. For class CII, however, the precise relationship between our approach and the previous one is not clear to us. In both approaches, one starts by enclosing the putative nodal loop in an enclosing loop. In our approach, one simply computes the winding number. In the previous approach, one has to choose the enclosing loop to be centrosymmetric, which perforce encloses both members of a pair of nodal loops related by ${\cal T}$.  One then has to extend the Hamiltonian in two extra dimensions in such a way that it remains gapped in the three-dimensional torus formed by the enclosing loop and the two extra dimensions, and then compute the invariant. We are unable to do this for generic centrosymmetric loops. A special loop for which we can perform the computation in presented in the appendix, for a model in which the nodal loops are topologically protected according to our classification. Unfortunately, both our winding number and the topological invariant of the previous approach are trivial on this centrosymmetric loop. The relation between the two approaches remains an important open question for class CII.

Our fourth result concerns the degeneracies of the zero-energy drumhead modes predicted to occur on an open surface in any nodal loop semimetal. The winding number is once again invaluable in determining whether zero-energy states (drumhead modes) exist on the surface at a given ${\bf k}$ in the surface Brillouin zone, and if so, what their degeneracies are. If loops with the same winding number have projections onto the surface Brillouin zone that overlap in some region, the drumhead modes will be doubly degenerate in that region. We show an explicit lattice example of this in class CII. By contrast, if loops with opposite winding numbers overlap, there will be no zero-energy modes in the overlap region, as we show with explicit examples in classes CI and DIII.

There are several possible platforms where the physics of topological nodal lines may be realized/studied. The compounds ${\mathrm{Ce}\mathrm{Pt}}_{3}\mathrm{Si}$ \cite{Bauer_2004}, ${\mathrm{Li}}_{2}{\mathrm{Pt}}_{3}\mathrm{B}$ \cite{Yuan_2006}, $\mathrm{Bi}\mathrm{Pd}$ \cite{Mondal2012} appear to be candidate materials for class DIII or AIII \cite{Matsuura_2013}. TNLSMs might also be realized in materials where there is a chiral symmetry or at least an effective chiral symmetry (e.g., $\mathrm{SrIrO}_3$ \cite{Chen_2015}). More recently, there has been a proposal for graphene networks whose band structure might realize  nodal lines \cite{Chen_2015_2}. A relatively new, but promising, platform has been proposed in driven systems \cite{Martin2017, Yang2018, Ozawa2019TopologicalQM}. In this platform, many topological systems have been modelled successfully both experimentally \cite{Esslinger_QHE_2023} and theoretically \cite{Bhat2021OutOE,Long2021}. Topological nodal lines may also be realized in the optical lattices of  ultracold atoms  \cite{Song_2019,Schafer2020}.

Coming to open questions, the relation between our winding number approach and the $Z_2$ topological invariant on centrosymmetric loops proposed in Refs. \onlinecite{Zhao_Wang_2013,Matsuura_2013,Zhao_Wang_2014} is certainly a pressing issue. In particular it is important to understand whether  a model which shows a nontrivial index in one approach is guaranteed to show nontriviality in the other. 

The effects of orbital magnetic fields on the spectra of semimetals is another area in which there are open questions. 
There has been quite a bit of study about Weyl semimetals in a magnetic field
\cite{Gooth_2023,Abdulla_2022}, and that produces a rich array of physics. The effect of magnetic fields 
on  nodal line
semimetals \cite{Rhim2015,Alberto2018,Li2018,Yang2018_2,Molina2018,Chen2022} have been studied for some simple cases. However, the generic case of nonplanar loops has not been studied.

Finally, a physically important question is the stability of topological semimetals in the presence of disorder and/or  interactions \cite{Fradkin_1986_I, Fradkin_1986, Goswami_etal_2011, Altland_2015,Roy_etal_2018, Kobayashi_etal_2014,Sbierski_etal_2014,Louvet_Carpentier_Fedoreko_2016,Rahul_2014, Pixley1_2016, Pixley2_2016, Pixley3_2017}. Some studies have been carried out for topological nodal loop semimetals. For chiral symmetry-respecting disorder, gapless modes are expected to be robust \cite{Matsuura_2013}. However, even if the disorder breaks the chiral symmetry, numerical work shows that surface states remain stable~\cite{Silva_2023} up to a critical disorder strength.  The situation might be different in the presence of long-range Coulomb interactions \cite{yejin_2016,Wang_Nandkishore_2017,PhysRevLett.124.136405}. We hope to address all these interesting questions for the various symmetry classes in greater generality in the near future.

\begin{acknowledgements}
The authors would like to thank Alexander Altland for very insightful discussions.  The authors also thank Andreas P. Schnyder for his insights.
FA would like to thank the Infosys Foundation for financial support and the ICTS for hospitality.
GM acknowledges the US-Israel Binational Science Foundation (grant no. 2016130), and the
hospitality of the Aspen Center for Physics (NSF grant no. PHY-1607611).  
A.D. was supported by the German-Israeli Foundation Grant No. I-1505-303.10/2019,
DFG MI 658/10-2, DFG RO 2247/11-1, DFG EG 96/13-1, CRC 183 (project C01),
and by the Minerva Foundation. A.D.
also thanks to the Israel Planning and Budgeting Committee (PBC) and the Weizmann Institute of
Science, the Dean of Faculty fellowship, and the Koshland Foundation for financial support.
The authors would like to thank the International Centre for Theoretical
Sciences (ICTS) for organizing the program -
Condensed Matter meets Quantum Information (code:
ICTS/COMQUI2023/9) and their hospitality where part of this work was completed.
\end{acknowledgements}

\appendix

\section{$\mathbb{Z}_2$ topological charge for nodal line in class CII on a specific loop}
\label{appsec:Z_2}
The definition of the $\mathbb{Z}_2$ invariant starts with a centrosymmetric loop in the physical 3D BZ. This loop is then extended in two extra momentum dimensions $\alpha,\beta$, with the Hamiltonian on this 3-Torus satisfying periodicity and having the symmetry of the CII class.  The charge is calculated on the 3-Torus using 
\cite{Zhao_Wang_2013,Matsuura_2013, Zhao_Wang_2014}
\begin{widetext}
\begin{align}\label{eq:CIIZ2}
    \nu_p = \frac{C_{p+1}}{2} \int_{S^{p-1}\times T^2} d\theta d\alpha d\beta ~ \epsilon^{\mu \nu \lambda} 
     Tr\left({\cal S} \tilde{G} \partial_{\mu} {\tilde G}^{-1} \tilde{G} \partial_{\nu} \tilde{G}^{-1} \tilde{G} \partial_{\lambda} \tilde{G}^{-1}\right)|_{\omega=0}
     ~ \mathrm{mod} ~ 2, 
\end{align}
\end{widetext}
where $p=2$ is the codimension of the Fermi surface (nodal line), ${\cal S}=\eta^z$ is 
the chiral operator and $\mu, \nu, \lambda \in (\theta, \alpha, \beta)$. The Green's 
function $\tilde{G}$ is given  by $\tilde{G}(\omega, {\bf k}) = 
\left(i\omega - \tilde{H}({\bf k}, \alpha, \beta)\right)^{-1}$.  The integration is done over 
$S^{p-1}\times T^2$, where $S^{p-1}=S^1$ is a centrosymmetric one dimensional  enclosing manifold (parametrized 
by $\theta$)
in the  three dimensional ${\bf k}$ space which keeps the antiunitary time-reversal 
and particle-hole symmetry of the Hamiltonian. The Hamiltonian $H({\bf k})$ 
is extended to $\tilde{H}({\bf k}, \alpha, \beta)$
maintaining time-reversal and charge conjugation symmetry: ${\cal T} \tilde{H}({\bf k}, \alpha, \beta){\cal T}^{-1} = 
\tilde{H}(-{\bf k}, -\alpha, -\beta)$, ${\cal C} \tilde{H}({\bf k}, \alpha, 
\beta){\cal C}^{-1} =-\tilde{H}(-{\bf k}, -\alpha, -\beta)$.  The extended Green's function 
$\tilde{G}(\omega, {\bf k})$ must satisfy the following two conditions 
\begin{subequations}
\begin{align} 
& \tilde{G}(\omega, {\bf k}, \alpha=0, \beta=0)|_{S^1} =  G(\omega, {\bf k})|_{S^1} \\
& \tilde{G}(\omega, {\bf k}, \alpha=\pi/2, \beta=\pi/2)|_{S^1} = (i\omega \pm \Delta)^{-1}|_{S^1},
\end{align}
\end{subequations}
where $\Delta$ is a constant. The above conditions ensure that a 1D topological insulator 
described by the Hamiltonian $\tilde{H}({\bf k}, \alpha=0, \beta=0) = H({\bf k})$ restricted  
on $S^1$, goes to a trivial state with an energy gap $\Delta$ at $ \alpha=\pi/2, \beta=\pi/2$. Furthermore, since $\alpha,\beta\in[-\frac{\pi}{2},\frac{\pi}{2})$, the Hamiltonian must be periodic in these two extra dimensions upto to unitary equivalence. 

Ideally, one would calculate this invariant for all possible centrosymmetric 1D enclosing loops extended into 3-Tori in the manner described above. If the invariant on any one such enclosing loops is nontrivial, the model is in a nontrivial nodal loop semimetal phase. A trivial result on a particular loop does not imply that the model is in a trivial phase.

\begin{figure*}[ht]
     \centering
     \includegraphics[width=0.9\linewidth]{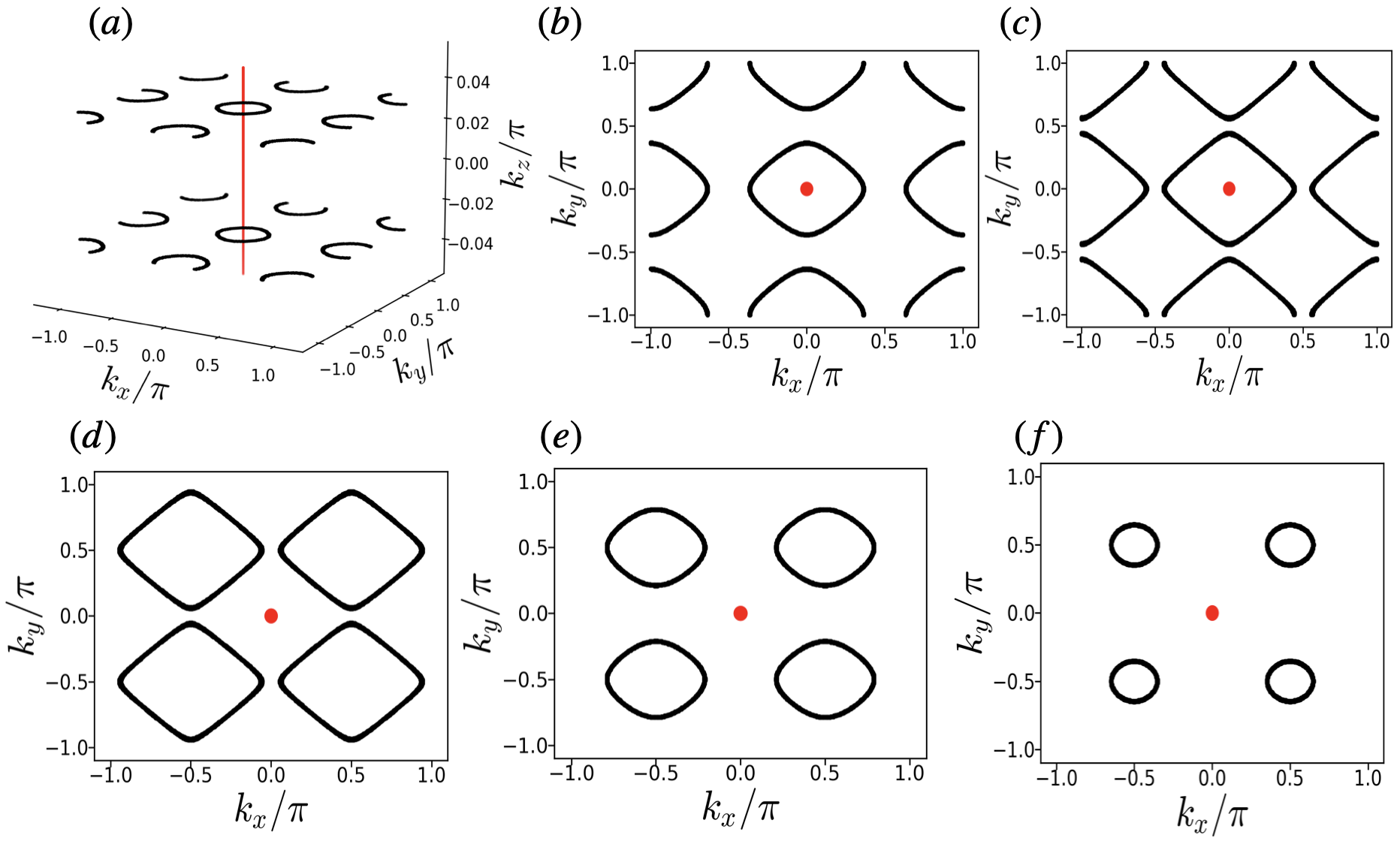}
     \caption{Evolution of the nodal loops with the (extension) parameters $\alpha$ and 
     $\beta$ to demonstrate that the extended Hamiltonian on enclosing red loop, passing 
     through the origin along the $k_z$ axis, always remains gapped. These nodal loops 
     are the solutions of the two \cref{eq:coskz2,eq:sinxy2} for different 
     values of $g = \cos^2{\alpha}\cos^2{\beta}$. In the figures from (a)-(f), the values 
     of $g = 1.0, 0.5, 0.42, 0.4, 0.3$ and $0.23$ respectively. Note that $g=1$ corresponds 
     to the gapless solution of  $ H_{\text{CII}, \text{Lat}}({\bf k})$. The system is 
     fully gapped for $g \lesssim 0.20$. In figures from 
     (b)-(f), we show solution only at  $k_z = k_{z0}$ plane. The solution at the time 
     reversed-momentum $k_z= -k_{z0}$ plane is identical to the solution at $k_z = k_{z0}$ plane.
     The red dot represents the origin $k_x=k_y=0$. } 
\label{fig:LoopEvol}
\end{figure*}

We are unable to compute the invariant on arbitrary centrosymmetric loops, because the extension is difficult to carry out in such a way that it satisfies all the required conditions.  We have been able to compute  this $Z_2$ invariant on a specific loop, the $k_z$ axis of the BZ. The calculation described in this subsection, while yielding a trivial result in a model we know to be nontrivial, will illustrate some of the difficulties that have to be overcome. We consider the Hamiltonian  $H_{\text{CII}, \epsilon}$ in \cref{eq:HCIIVe}
\begin{align}
H_{\text{CII}, \epsilon}({\bf k}) = H_0({\bf k}) + A \eta^y\otimes a_3 + V_{\epsilon} \eta^x \otimes \epsilon,
\end{align}
which has a robust gapless NLSM phase, as demonstrated through a simple gapless solution in \cref{eq:CIIfinalCond}. For momentum independent $M, A$, $V_{\epsilon}$, we have a pair of nodal loops lying on the $k_x$-$k_y$ plane at a fixed $k_z$ values $\pm m V_{\epsilon}/A$. We choose the simplest centrosymmetric enclosing loop, the $k_z$ axis. Making this choice means one has to consider large values of $k_z$,  beyond the validity of the continuum model we made near the $\Gamma$ point. To regulate this, and more importantly, to make the model applicable to crystalline solids, we go to a lattice model with the same continuum limit.  A lattice model can straightforwardly be obtained from $H_{\text{CII}, \epsilon}({\bf k})$  by replacing  ${\bf k}$ by $\sin{{\bf k}}$ and ${\bf k}^2$ by $\cos{\bf k}$, after which we obtain 
\begin{align}
    H_{\text{CII}, \text{Lat}}({\bf k}) =  \sum_{i=1}^{3} \sin{k_i} \Gamma_i + m_4({\bf k})\Gamma_4 \nonumber\\
    + A \eta^y\otimes a_3  + V_{\epsilon} \eta^x \otimes \epsilon,
\end{align}
where we have chosen a specific momentum dependence in $m_4({\bf k}) = m - t \cos{k_z}$ for reasons that will become clear shortly. The Gamma matrices remain as before; $\Gamma_1= \eta^x \tau^z \sigma^x$,  $\Gamma_2= \eta^x
\tau^z \sigma^y$,  $\Gamma_3= \eta^x \tau^y \sigma^0$ and  $\Gamma_4= \eta^x \tau^x \sigma^0$. 
The gapless solutions of $H_{\text{CII}, \text{Lat}}({\bf k})$ can be obtained 
from the \cref{eq:CIIfinalCond} by replacing $k_i$ by $\sin{k_i}$:
\begin{subequations}
\begin{align}\label{eq:latticeGC}
    &  \sin^2{k_x} + \sin^2{k_y} + \sin^2{k_z} + m_4^2({\bf k})= A^2+V_{\epsilon}^2 \\
    &  A \sin{k_z} =\pm V_{\epsilon}m_4({\bf k}).
\end{align}
\end{subequations}
Simplifying the second condition, we find that the nodal loops lie parallel to  $k_x$-$k_y$ plane at 
two fixed $k_z=\pm k_{z0}$ values determined by  
\begin{align}\label{eq:coskz}
    \cos{k_{z0}} = \frac{m t V^2_{\epsilon} \pm \sqrt{\left(mtV^2_{\epsilon}\right)^2 - (V^2_{\epsilon}t^2 + A^2) 
     (m^2 V^2_{\epsilon} -A^2)}}{A^2 + V^2_{\epsilon} t^2}, 
\end{align}
The first condition can be rewritten as 
\begin{align}\label{eq:sinxy}
   \sum_{i=x,y} \sin^2{k_i}  = V_{\epsilon}^2\left(1 - \frac{m^2_4({k_{z0}})}{A^2}\right) 
    + (A^2 - m_4^2(k_{z0})). 
\end{align}
The nodal loops exist if both conditions have simultaneous solutions. From   \cref{eq:coskz} we see that there are are potentially four values $\pm k_{z0},\ \pm k_{z0}'$. However, we can choose our parameters $(m, t, A, V_{\epsilon})$ in such a way that 
the above two conditions allow only two solutions. This is necessary 
because the enclosing loop (which is the $k_z$ axis) should not enclose more than two 
nodal loops to get a nontrivial $Z_2$ invariant $\nu_p$. We work with the following choice 
of parameter values: $m=0.5$, $t=-1.4$, $A=2.0$, $V_{\epsilon}=0.1$. Solving \cref{eq:coskz}, we get $\cos{k_{z0}}\approx 0.995$ and $-0.999$ for 
the plus and minus signs, respectively. Substituting these values into the second condition, \cref{eq:sinxy}, we obtain $\sin^2{k_x} + \sin^2{k_y} \approx 0.414$ and 
$3.20$ for the plus and minus signs, respectively. Clearly, 
the second condition has no solution for $\cos{k_{z0}}\approx -0.999$. The full set of nodal loops is depicted in \cref{fig:LoopEvol}(a). Each loop has a nonzero winding number, so the model is a topologically protected nodal loop semimetal.

Now we are ready two make a two-parameter extension. We consider the following extension of 
$ H_{\text{CII}, \text{Lat}}({\bf k})$ 
\begin{widetext}
\begin{align}
    \tilde{H}_{\text{CII}, \text{Lat}}({\bf k}, \alpha, \beta) = \left( \sum_{i=1}^{3} \sin{k_i} \Gamma_i + 
    m_4({\bf k})\Gamma_4   + V_{\epsilon} \eta^x \otimes \epsilon \right)\cos{\alpha}\cos{\beta} + 
    A \eta^y\otimes a_3 + m_5({\bf k})\sin{\alpha}\cos{\beta} \Gamma_5 + m_6({\bf k}) \sin{\beta}\Gamma_6
\end{align}
\end{widetext}
where $m_5({\bf k}) = m_5(-{\bf k})$ and $m_6({\bf k}) = m_6(-{\bf k})$ are even functions of the
momenta.  The two other $\Gamma$ matrices are $\Gamma_5 = \eta^x \tau^z \sigma^z$, $\Gamma_6 = \eta^y$, which anticommute with $\Gamma_i$, 
$i=1, 2, 3, 4$. Note that both $\Gamma_5$ and  $\Gamma_6$ are odd under time reversal and charge conjugation such that $\tilde{H}_{\text{CII}, \text{Lat}}({\bf k}, \alpha, \beta)$ obeys: ${\cal T}  \tilde{H}_{\text{CII}, \text{Lat}}({\bf k},\alpha, \beta){\cal T}^{-1} = \tilde{H}_{\text{CII}, \text{Lat}}(-{\bf k}, -\alpha, -\beta)$, and  ${\cal C} \tilde{H}_{\text{CII},  \text{Lat}}({\bf k}, \alpha, \beta){\cal C}^{-1} =-\tilde{H}_{\text{CII}, \text{Lat}}(-{\bf k}, -\alpha, -\beta)$. Both parameters lie in the range $[-\frac{\pi}{2},\frac{\pi}{2})$. To make sure that the Hamiltonian lies on a 2-Torus in $\alpha, \beta$ space, we need to ensure that $H({\bf k},\alpha,\pi/2)$ is unitarily equivalent to $H({\bf k},\alpha,-\pi/2)$, and also that $H({\bf k},\pi/2,\beta)$ is unitarily equivalent to $H({\bf k},-\pi/2,\beta)$. It can be easily checked that this is true with the unitary transformation being none other than $\eta^z$.

The condition that the extended Hamiltonian remains gapped on $S^1\times T^2$ can be rephrased as saying that as we 
change $\alpha$ and $\beta$ in their entire range, the nodal loops never hit the enclosing loop along the $k_z$ axis. The zero energy conditions are straightforward generalizations of \cref{eq:Hsolution}.
\begin{align}\label{eq:coskz2}
    \cos{k_{z0}} = \frac{m t V^2_{\epsilon} \pm \sqrt{\left(mtV^2_{\epsilon}\right)^2 - (V^2_{\epsilon}t^2 + A^2) 
     (m^2 V^2_{\epsilon} -A^2)}}{A^2 + V^2_{\epsilon} t^2}, 
\end{align}
which is identical to \cref{eq:coskz} and 
\begin{align}\label{eq:sinxy2}
    \sum_{i=x,y} \sin^2{k_i} = V_{\epsilon}^2\left(1 - \frac{m^2_4({k_{z0}})}{A^2}\right) 
    + \frac{A^2 - m_4^2(k_{z0})}{\cos^2{\alpha}\cos^2{\beta}}, 
\end{align}
which is again almost identical to \cref{eq:sinxy}, except the factor $\cos^2{\alpha}\cos^2{\beta}$ 
in the last term. The above two conditions are obtained by assuming  $m_4({\bf k}) = m_5({\bf k}) = m_6({\bf k})$, 
and $m_4({\bf k}) = m - t\cos{k_z}$. Now the key point is that if $A-m_4(k_{z0})>0$ , then  
the right hand side of \cref{eq:sinxy2} always remains positive for arbitrary values of $\alpha$, $\beta$. This ensures that the extended Hamiltonian on the enclosing loop (times the 2-Torus) is always gapped. 
For the choice of parameters we have made, we get $A-m_4(k_{z0}) = 0.1$ which satisfies the 
crucial condition $A-m_4(k_{z0})>0$. 
Clearly, at $\alpha= \beta = \pi/2$, the extended Hamiltonian is fully gapped everywhere in $k$-space.  
For the sake of illustration, we have shown how the nodal loops evolve as we vary the quantity $g=\cos^2{\alpha}\cos^2{\beta}$, through a few plots in \cref{fig:LoopEvol}. Qualitatively, the loops maintain the same $k_z$, expand, and merge with loops starting near the face centers of the zone at a Lifshitz transition at some value of $g=\cos^2{\alpha}\cos^2{\beta}$. After the merger, as $g$ is further reduced, the loops shrink, and finally disappear for $g=0$, leaving behind a trivial insulator.

This completes the description of the two-parameter extension of the Hamiltonian. It is now straightforward to 
numerically compute the integral in \cref{eq:CIIZ2} on  the enclosing loop along
the $k_z$ axis from $k_z=-\pi$ to $\pi$. The integral is well-defined everywhere in the domain 
of integration, and gives a trivial $Z_2$ invariant. As mentioned before, the winding number of each nodal loop is nonzero, and the model itself is nontrivial. We note that our winding number computed on the $k_z$-axis also vanishes. The model is nontrivial because the winding number computed on a line parallel to the $k_z$ axis lying outside the nodal loops (at $k_x=k_y=\pi/2$, say) gives a nonzero answer.  However, since this enclosing loop is not centrosymmetric, we are not able to compute the 3-Torus invariant on it.  It could be that there are other centrosymmetric loops in this model which do produce a nontrivial 3-Torus invariant. However, as illustrated in this subsection, constructing an extension of the Hamiltonian to the 3-Torus satisfying all the required conditions is highly nontrivial.



\section{class AII in 3D}
\label{App:AII}
The classification by Refs. \cite{Zhao_Wang_2013,Matsuura_2013} predicts a nontrivial index 
for nodal lines in class AII in three dimensions. In the following, we show that there cannot be a stable nodal loop phase  in the minimal model in  class AII in three dimensions. However, class AII does have a generic, stable Weyl semimetal phase.

The minimal dimension of the Dirac Hamiltonian in class AII  (which has time-reversal 
symmetry ${\cal T}^2 =-1$ only) in three spatial dimension is four. The transition from 
a topological to a trivial  insulating states can be described by  a  four by four Dirac 
Hamiltonian, 
\begin{align}
H_0({\bf k}) = \sum_{i=1}^{3} k_i \gamma_i  + m \gamma_4,  
\label{eq:HAII}
\end{align}
where the Gamma matrices  $\gamma_{\mu}$, $\mu = 1, 2, .., 4$, satisfy the usual 
anticommutation relation $\{\gamma_{\mu}, \gamma_{\nu}\} = 2\delta_{\mu \nu}$. The 
fifth Gamma matrix is $\gamma_5 = \gamma_1 \gamma_2 \gamma_3 \gamma_4.$ 
The full space of Hermitian four by four Hamiltonians is spanned by including the
identity $\gamma_0 = \mathbb{I}_4$ and another ten matrices $\gamma_{ab} = 
-\frac{i}{2} [\gamma_a, \gamma_b]$,  where $a, b =1, 2, ...., 5$, and $a<b$.  
In  what follows, it is convenient to work with a given representation of the Gamma matrices. 
Our chosen representation is the following: $\gamma_1=\tau_z \sigma_x, ~  \gamma_2=\tau_z 
\sigma_y, ~\gamma_3=\tau_y \sigma_0, ~\gamma_4=\tau_x\sigma_0$, and 
$\gamma_5= \tau_z\sigma_z$, where $\tau$, $\sigma$ act on the orbital and spin 
space respectively.  
The Dirac Hamiltonian $H_0({\bf k})$ is symmetric under the time-reversal which is  
realized by ${\cal T}=-i\sigma_y K$ , ${\cal T}^2=-1$.

In the spirit of breaking all accidental lattice  symmetries except translations, we first look for such symmetries in the Hamiltonian of \cref{eq:HAII}. Since the matrix $\gamma_5$  is absent 
in $H_0({\bf k})$, the Hamiltonian is also symmetric under chirality $ {\cal S} H_0({\bf k}) 
{\cal S}^{-1} = - H_0({\bf k})$, with ${\cal S} = \gamma_5$. To obtain a generic Hamiltonian in 
class AII from $H_0({\bf k})$, we must add terms which break this chiral symmetry 
but preserve the time-reversal  symmetry. The transformation properties of all the  
Dirac matrices  are given in \cref{tab:table4}. 

\begin{table}[ht]
  \begin{center}
    \begin{tabular}{p{40mm} p{20mm} p{8mm}} 
      \hline 
      \hline
      \textbf{Operators} & ${\cal T}$ & ${\cal S}$ \\ [1mm]
      \hline
     $\gamma_i$, ~ $(i=1, 2, 3)$ & $-1$ & $-1$ \\ [1mm]
     $ \gamma_4$  & $+1$ & $-1$ \\ [1mm]
     $ \gamma_5$  & $-1$  & $+1$ \\ [1mm]
     $ \gamma_0$  & $+1$ & $+1$ \\ [1mm]
     $ \epsilon = \gamma_{45}$  & $+1$ & $-1$ \\ [1mm]
     $ {\bf b} = (\gamma_{23}, \gamma_{13}, \gamma_{12}) $  & $-1$ & $+1$ \\ [1mm]
     $ {\bf p} = (\gamma_{14}, \gamma_{24}, \gamma_{34}) $  & $+1$ & $+1$ \\ [1mm]
     $ {\bf b}' = (\gamma_{15}, \gamma_{25}, \gamma_{35})$  & $-1$ & $-1$ \\ [1mm]
      \hline 
    \end{tabular}
    \caption{Transformation properties of Dirac matrices under time reversal ${\cal T}$ and 
     chirality ${\cal S} = \gamma_5$.         \label{tab:table4}
}
  \end{center}
\end{table}

From the \cref{tab:table4}, we see that only the terms $\gamma_0 = \mathbb{I}_4$, $\epsilon = \gamma_{45}$,  
and ${\bf p} = (\gamma_{14}, \gamma_{24}, \gamma_{34})$ are allowed 
 without momentum dependence (or with quadratic momentum dependence). Therefore the most general 
Hamiltonian in class AII is 
\begin{align}\label{eq:HAII}
H_{\text{AII}} ({\bf k}) = H_0({\bf k}) + {\bf V}_{p} \cdot {\bf p} + V_{\epsilon} \epsilon. 
\end{align}
Clearly, the term  ${\bf V}_{p} \cdot {\bf p}$ breaks the chiral symmetry and  makes sure 
that the Hamiltonian $H_{\text{AII}} ({\bf k})$ has only the symmetries required by class AII. Note that ${\bf V}_{p}$
and $V_{\epsilon}$ can be any even function of momenta without altering the time-reversal 
symmetry of  $H_{\text{AII}} ({\bf k})$. A minimal choice would be $m\to m({\bf k})= m_0 - 
\sum_{ij} t^{(m)}_{ij} k_i k_j$, ~ ${\bf V}_p \to {\bf V}_p({\bf k})= {\bf V}_{p0} - \sum_{ij}
 t^{({\bf V}_p)}_{ij} k_i k_j$, $V_{\epsilon} \to V_{\epsilon} ({\bf k})= V_{\epsilon 0} - 
\sum_{ij} t^{(V_{\epsilon} )}_{ij} k_i k_j$ where $m_0, {\bf V}_{p0}, V_{\epsilon 0},   
t^{(m)}_{ij}$,  $t^{({\bf V}_p)}_{ij}$ and  $t^{(V_{\epsilon})}_{ij}$  are all real. For 
a generic choice of  $t^{()}_{ij}$, it is easily checked that all the crystalline symmetries
are broken. 
 
 We want to know whether  $H_{\text{AII}} ({\bf k})$ possesses a stable gapless 
 phase. If it does,  then we would like to know the nature of the gapless phase (e.g. Weyl semimetal, 
 nodal loop semimetal, etc.). The Hamiltonian $H_{\text{AII}} ({\bf k})$ can be easily diagonalized 
 to obtain   the following energy spectrum
 \begin{align}
 E^2({\bf k}) = {\bf k}^2 + m^2 +  {\bf V}_p^2 + V_{\epsilon}^2 \pm 2\sqrt{V_{\epsilon}^2 {\bf k}^2 +  
 {\bf V}^2_p {\bf k}^2 - |{\bf V}_p \cdot {\bf k}|^2 }. 
 \end{align}
The locus of  gapless solutions satisfies the following condition
\begin{align}
{\bf k}^2 + m^2 + {\bf V}_p^2 + V_{\epsilon}^2 =  2\sqrt{V_{\epsilon}^2 {\bf k}^2 +  
 {\bf V}^2_p {\bf k}^2 - |{\bf V}_p \cdot {\bf k}|^2 }.
\end{align}
This condition can be re-written as 
\begin{align}
\left({\bf k}^2 - {\bf V}_p^2 - V_{\epsilon}^2 \right)^2 + m^2 \left(m^2 + 2{\bf k}^2 
+ 2{\bf V}_p^2 + 2V_{\epsilon}^2\right)\nonumber\\
+ |{\bf V}_p \cdot {\bf k}|^2 = 0. 
\end{align}
Clearly, for a nontrivial solution, we must have i) $ {\bf k}^2 = {\bf V}_p^2 + V_{\epsilon}^2$,  ii) $m = 0$, and 
iii) ${\bf V}_p \cdot {\bf k} = 0$. The intersection of three 
surfaces, if it occurs at all, must generically describe isolated points. A small change in the parameters of the Hamiltonian cannot 
remove these solutions. Therefore  $H_{\text{AII}} ({\bf k})$ does possess a stable gapless 
phase, which is a topological Weyl semimetal. This is another way to view the generality of the Murakami construction \cite{Murakami2007}. Recall that Murakami \cite{Murakami2007} broke inversion but maintained time-reversal symmetry, putting the model in class AII. The isolated Weyl points can be enclosed in a non-centrosymmetric 2D surface. The Hamiltonian restricted to this surface is in class A, and its topological index is the Chern number of the Weyl point.

\bibliographystyle{apsrev}
\bibliography{export}

\end{document}